# *EINSTEIN, SCIASCIA, MAJORANA, AMALDI* E

# IL RAPPORTO TRA INTELLETTUALI E POTERE

Erasmo Umberto M. RECAMI, Umberto Victor G. RECAMI, e Erasmo RECAMI (*)

[***Abstract:*** *see the next page*]

*Compito supremo dello scienziato è di pervenire a quelle leggi elementari universali partendo dalle quali il cosmo può essere costruito con la pura deduzione. Lo conduce l'"intuizione, e il suo sforzo quotidiano scaturisce direttamente dal cuore. (Albert Einstein)*

*Voglio soltanto fare presente che per me l'espressione "rifiuto della scienza" vale "rifiuto della scienza a un certo punto di fronte a certe ricerche, a certe scoperte". E cioè rifiuto da parte degli scienziati stessi. (Leonardo Sciascia)*

———————————

(*) *Fac. di Ing., Università degli studi di Bergamo, Bergamo, Italy,*
    and *INFN-Sezione di Milano*, *Milan, Italy.*

*Abstract (in English)* -- In Europe (in Italy, for example), and in the States, the opinion is widely spreading that the negative consequences of modern progress are the fault of "Science" (we'd say, *then*, of the scientists, or, rather, of the intellectuals). It is very important that both scholars of the "Two Cultures" try to slow down such a fall into a Second Middle Age… A lively debate on this topic took place among the famous writer Leonardo Sciascia and Italian physicists like Edoardo Amaldi and Emilio Segré, when Sciascia wrote his known book about the disappearance of Ettore Majorana, a book that rather arbitrarily presented Majorana (a very reserved theoretical physicist, indeed) as the example of the scientist that keeps his discoveries secret when afraid of possible evil applications. We wish to contribute now to such a question [we already contributed to that debate, also because of our position, at that time, of good friends of both L.Sciascia and E.Amaldi et al.], since many special meetings did recently return to it, while celebrating the centenary of Majorana's birth (2006), or 30 years (2005) from Sciascia's book publication. It appeared natural to us to start with the figure of Einstein, who, even if extremely peaceable, supported the atomic bomb construction by his letter to Roosvelt [whose original is here reproduced, and translated into Italian]. We discuss first the significance of part of Einsteins's scientific work (which flourished in particular one century before the recent year 2005): We seize this opportunity for presenting a derivation of the "twins paradox", so simple that it has been easily taught to Senior High School last-year students or University first-year undergraduates; all the present work, more in general, is addressed mainly to them. In the second part of this paper, we analyse the general meaning of Einstein's contributions to philosophy and pedagogy. The third part is entirely devoted to our *main subject*, namely, to discussing the "Responsibility of the Intellectuals". The reader interested chiefly in this last part, can find it re-expounded in a self-contained way in the *Appendix*. More information can be found by checking, below, the Contents ("Indice") of the present article. The occasion for writing this paper (in Italian) was a wish to help young Italian students.

# INDICE:

# PARTE PRIMA

## L'OPERA DI EINSTEIN E SUOI COLLEGAMENTI CON LA CULTURA IN SENSO LATO

*Introduzione, e Progetto di questo scritto:*
[Premessa formale: Il materiale qui contenuto non può essere riprodotto senza la citazione della fonte, e/o senza il permesso degli eventuali detentori di copyright].

Cent'anni fa **Albert Einstein** –oscuro impiegato in un Ufficio Brevetti di Berna (non essendo stato accettato come assistente da alcuna università)-- si rivelava al mondo della scienza all'età di 26 anni, pubblicando durante un solo anno quattro articoli, ognuno dei quali gli avrebbe meritato un premio Nobel. Il primo di essi uscì il 17 marzo 1905. L'anno *2005 è stato dichiarato Anno Internazionale della Fisica* proprio per celebrare il centenario di quell' *annus mirabilis,* il 1905.

Ma misurare il contributo di Einstein alla scienza e alla nostra visione del mondo mettendo in fila le sue pur numerose scoperte è riduttivo. Sono i concetti stessi di spazio, tempo, realtà, cosmo che dopo Einstein –soprattutto dopo le sue teorie "Relativistiche"-- hanno acquistato un significato nuovo. Ciò porta naturalmente a individuare legami della sua opera con quella di filosofi come Immanuel Kant**,** di letterati come Luigi Pirandello, e di movimenti artistici come quello del Cubismo. Ricordiamo inoltre che, partendo dalla nota relazione einsteiniana *$E=mc^2$*, si arriva alla interpretazione dei processi di fusione termonucleare che la natura usa *al centro delle Stelle* (per esempio, del nostro Sole) per produrre energia.

La coscienza einsteiniana che il sapere è opera di molti, nel senso che le sue stesse teorie saranno sostituite da altre più generali, lo avvicina poi ad autori classici come Seneca. Perché Einstein, come Seneca (benché non quanto lui), fu costretto ad avere dei rapporti ambigui col potere.

Einstein, poi, con la sua *lettera del 2 agosto 1939 al Presidente degli Stati Uniti* –che vedremo in seguito (nel testo originale, e *nella nostra traduzione in italiano*)  convinse Roosvelt a fare partire il "Progetto Manhattam"  per la costruzione della bomba atomica, mettendo così in  eclatante evidenza il problema della responsabilità degli scienziati (trattato in tempi recenti, in Italia, soprattutto da **Leonardo Sciascia** nel suo saggio su "*La scomparsa di Majorana*") e, più in generale, del delicato rapporto tra intellettuali e potere.

Infine, la fuga di Einstein e di altri (come Leo Slizard, che convinse Einstein a scrivere la suddetta lettera a Roosvelt) dalla Germania di Hitler, e lo stesso progetto Manhattam --così come alcuni documenti riservati sugli scienziati tedeschi recentemente resi pubblici dal Servizio Segreto britannico--, richiederebbero un esame storico sugli anni dal 1930 al 1945 (dalla nascita del Nazismo in Germania, allo scoppio delle bombe A sul Giappone).  La tragedia della II Guerra Mondiale, e delle bombe su Hiroshima e Nagasaki, permette collegamenti anche con autori (tra quelli di lingua madre inglese) come Samuel Beckett**,** che da tali tragedie furono profondamente influenzati nella loro opera.

**Cominciamo comunque –prima di passare a ciò che più ci interessa-- con l'accennare *brevemente* ai contributi di Einstein alla Meccanica Quantistica e alla fisica molecolare.**

*Cenno al contributo di A.E. alla Fisica molecolare e atomica:*
Per comprendere quanto innovatrice fu l'opera di Einstein anche in questi settori, per noi marginali, bisogna ricordare che all'inizio del secolo appena trascorso moltissimi scienziati *non* credevano ancora alla reale esistenza di atomi e molecole. Einstein allora, nel 1904, si propose di individuare un fenomeno che potesse sperimentalmente e chiaramente mettere in mostra la loro esistenza e nel contempo fornire indicazioni sulle loro dimensioni. Scoprì che un corpicciolo sufficientemente minuscolo (ma pur sempre visibile al microscopio) immerso in un liquido avrebbe dovuto apparire soggetto a un caratteristico moto "a zig zag", a causa dei suoi urti con le molecole del liquido: ovvero, a causa dell'agitazione termica delle molecole. Einstein non sapeva che tale fenomeno --*il moto browniano*—fosse già stato osservato sperimentalmente; egli però ne fornì la derivazione teorica e l'interpretazione dettagliata, e *ciò* permise tra l'altro di valutare le dimensioni delle molecole. Questo suo lavoro convinse praticamente tutti della effettiva esistenza di molecole e atomi.

Ancora: una dozzina d'anni più tardi Einstein introdusse, in base a considerazioni eleganti e generali, il processo di "emissione stimolata", che sta alla base del *laser*. Einstein, quindi, può essere considerato anche il padre della luce laser.

La fisica atomica e molecolare e la fisica del laser hanno successivamente compiuto passi da gigante, che sono sotto gli occhi di ognuno. Ma non tutte le strade aperte da Einstein --come cercheremo di indicare-- sono poi andate avanti secondo le sue intenzioni.

*Cenno al contributo di A.E. alla Meccanica Quantistica:*
Un articolo del 1905, di cui stiamo per dire, gli procurerà effettivamente (e soltanto nel 1922) un premio Nobel; ma esso non riguarda la Relatività Speciale: dopo tanti anni l'Accademia non si fidava ancora dei suoi contributi più rivoluzionari, quelli che a noi qui più interessano...

Einstein riprese l'ipotesi quantistica (formulata nel 1900 da Planck, ma rimasta senza molto seguito in quei cinque anni, perché nessuno, nemmeno Planck, aveva avuto il coraggio di accettarla davvero), dimostrando che la luce è costituita da "granuli": i fotoni. E così mise in moto l'edificazione della Meccanica Quantistica (MQ), edificazione a cui contribuì decisamente anche molto più avanti.

Einstein contribuì dunque in maniera sostanziale alla costruzione della MQ; ma poi la considerò solo come un prodotto intermedio e incompleto, allontanandosi (forse a ragione) dalla stragrande maggioranza degli altri fisici.

**Passiamo ai contributi di Einstein di maggiore rilevanza per i concetti di spazio e di tempo (e di spazio-tempo), e per la cosmologia (nel quale settore, coi modelli di cosmo che le sue equazioni permettono di costruire, fece per esempio passare il problema cosmologico dal puro settore filosofico a quello scientifico-filosofico e scientifico).**

*Relatività e "Assolutività":*
Uno degli eterni dilemmi della scienza sta nella scelta tra una descrizione continua e una descrizione discreta della natura fisica. Einstein, pur avendo tanto contribuito alla nascita della Meccanica Quantistica (che si rifà al ***discreto***), optò decisamente per il ***continuo*** nelle sue teorie relativistiche. Cominciamo con la Relatività Speciale, pure pubblicata nell'anno favoloso 1905, all'età di ventisei anni.

Prima di Einstein i fisici avevano cominciato ad accorgersi che le misure della distanza spaziale e della distanza temporale tra due *eventi* A e B erano probabilmente non assolute, ma **relative** all'osservatore. Dalla Relatività Speciale (RS) si deduce subito, però, che **è assoluta** *la distanza spazio-temporale, quadridimensionale,* tra i due eventi A e B. La RS quindi ci insegna a costruire quantità assolute a partire da quantità relative: essa avrebbe ben potuto chiamarsi "Teoria della Assolutività"... Nel 1908 Minkowski (il collega di Einstein che ha dato il nome allo spazio-tempo) poté dichiarare di conseguenza:

*"D'ora in poi lo spazio di per sé e il tempo di per sé sono destinati ad affondare completamente nell'ombra, e soltanto lo spazio-tempo, una sorta di unione di entrambi, può conservare un'esistenza indipendente"*.

Apriamo una lunga parentesi su
***Invarianza della distanza spazio-temporale; e "Paradosso dei Gemelli":***
Abbiamo sopra accennato al fatto che già prima di Einstein i fisici si erano accorti che la velocità *c* della luce nel vuoto non dipendeva dalla velocità della sorgente S, ed era quindi la medesima (invariante) per tutti gli osservatori, qualunque fosse il loro moto rispetto alla sorgente [ovvero, sia che fossero fermi rispetto ad S, sia che ad S si stessero avvicinando, sia che da S si stessero allontanando – con qualsiasi velocità]. Si erano inoltre già accorti che, invece, dati due eventi $e_1$ ed $e_2$, la loro distanza spaziale $\Delta l$ e la loro distanza temporale $\Delta t$ non avevano valori univoci, ma dipendevano dall'osservatore (in Relatività galileana ed einsteiniana si considerano sempre e solo osservatori inerziali; inoltre col simbolo $\Delta l^2$ si indica, appunto, il quadrato $\Delta l^2 \equiv \Delta x^2 + \Delta y^2 + \Delta z^2$ della distanza spaziale).

Einstein ci ha però insegnato che è *assoluta*, cioè uguale per tutti gli osservatori, la distanza spazio-temporale (quadrimensionale) $\Delta s$ tra due eventi qualsiasi $e_1$ ed $e_2$, costruita *generalizzando* l'usuale teorema di Pitagora [cfr. Fig.1] per le quattro dimensioni $(t;x,y,z)$ [cfr. Fig.2]. Il fatto notevole è che in tale "formula di Pitagora generalizzata" la parte temporale $c^2\Delta t^2$ entra *con il segno opposto* a quello della distanza spaziale $\Delta l^2$, così che la distanza (quadrimensionale) invariante si scrive $\Delta s^2 \equiv \Delta l^2 - c^2\Delta t^2$ [il tempo va moltiplicato per la velocità $c$ per ovvie ragioni dimensionali, cioè per sommare non tempi a lunghezze (cosa proibita in fisica), ma lunghezze con lunghezze]. Anzi, passando alla convenzione sui segni attualmente più adottata, scriveremo la distanza spazio-temporale come $\Delta s^2 \equiv c^2\Delta t^2 - \Delta x^2 - \Delta y^2 - \Delta z^2$.

Come si diceva, Einstein ci ha insegnato che, dati due osservatori (e due eventi qualsivoglia $e_1$ ed $e_2$):

$$\Delta s_1^2 = \Delta s_2^2$$

dove $\Delta s_1$ e $\Delta s_2$ rappresentano la distanza spazio-temporale tra $e_1$ ed $e_2$ calcolata, rispettivamente, dal primo e dal secondo osservatore; *esse sono uguali.* La Relatività, ripetiamo, ci insegna a costruire quantità valide per ogni osservatore partendo da quantità relative agli osservatori. Sarebbe stato più corretto, quindi, chiamarla Teoria dell'Assolutività...; e Einstein inizialmente la chiamò infatti Teoria dell'Invarianza.

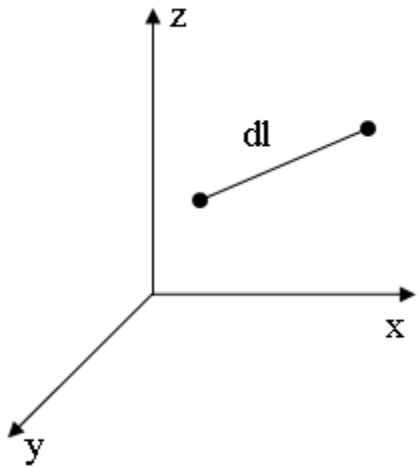

$$dl = \sqrt{dx^2 + dy^2 + dz^2}$$

$$dl^2 = dx^2 + dy^2 + dz^2$$

*Fig. 1-- Teorema di Pitagora nello spazio*

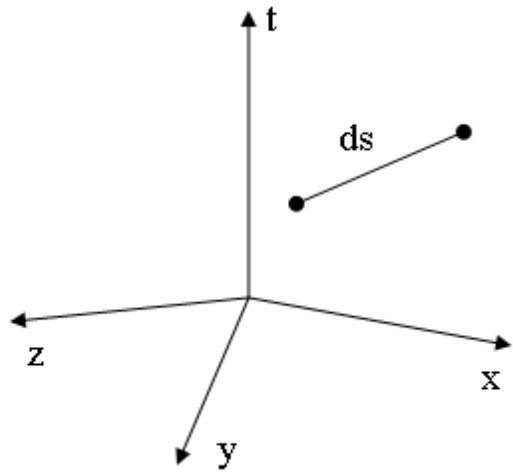

*Fig. 2-- Teorema di Pitagora generalizzato (nello spazio-tempo):  $-ds^2 \equiv dx^2 + dy^2 + dz^2 - c^2 dt^2$*

Il fatto che nella nostra definizione di $\Delta s^2$ la distanza temporale entri con un segno diverso da quello della distanza spaziale lo si può verificare nel semplice caso della luce. Se $e_1$ ed $e_2$ sono due eventi della vita *di una particella di luce* (la quale ha la medesima velocità per ogni osservatore), dato che lo spazio percorso in un certo tempo è ben noto essere *il prodotto* della velocità per quell'intervallo di tempo, si avrà che $c^2\Delta t_1^2 - \Delta l_1^2 = 0$ e analogamente che $c^2\Delta t_2^2 - \Delta l_2^2 = 0$ ; in tal caso si ha $\Delta s_1^2 = \Delta s_2^2$ (essendo entrambi nulli), proprio usando la definizione $\Delta s^2 \equiv c^2\Delta t^2 - \Delta l^2$, con *il meno* tra $c^2\Delta t^2$ e $\Delta l^2$ . L'uguaglianza $\Delta s_1^2 = \Delta s_2^2$ *vale però* non solo per la luce, ma per due eventi $e_1$ ed $e_2$ della vita di un oggetto *qualsiasi*.

Supponiamo ora che, nell'origine O (ossia, per semplicità, nel punto 0 al tempo 0), noi si venga raggiunti e superati da quello che riconosciamo come un nostro gemello, a bordo di un razzo dotato di velocità costante $v = X_0 / T_0$ (ved. Fig. 3). [Per facilitare le cose consideriamo, oltre al tempo *t*, solo un asse spaziale, *x* ].

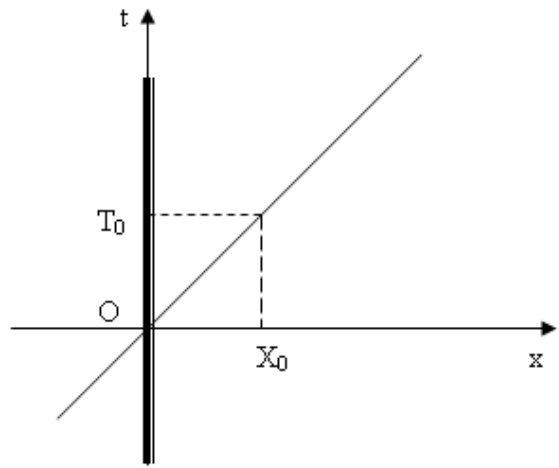

*Fig. 3*

Al passare del tempo, noi rispetto a noi stessi rimarremo sempre nel punto 0, così che la nostra "traiettoria" nello spazio-tempo (*t,x*) sarà l'asse verticale t. La "traiettoria" del nostro gemello sarà invece la retta inclinata (tali "traiettorie" si chiamano pomposamente "linee di universo"). Consideriamo due eventi lungo il cammino del gemello, e la distanza quadrimensionale che li separa, calcolata rispettivamente da noi e dal gemello. Attribuiamo alle nostre misure il pedice (indice al piede) 1; a quelle del nostro gemello è qui opportuno assegnare il pedice 0. Noi calcoleremo, al solito, $\Delta s_1^2 = c^2\Delta t_1^2 - \Delta x_1^2$ , che varrà $\Delta s_1^2 = c^2\Delta t_1^2 - v^2\Delta t_1^2$ , dato che rispetto a noi il gemello ha la velocità costante *v* (e pertanto, ovviamente, $\Delta x_1^2 = v^2\Delta t_1^2$ ). Il nostro gemello, invece calcolerà un $\Delta s_0^2 =_= c^2\Delta t_0^2 - \Delta x_0^2$ semplicemente uguale a $c^2\Delta t_0^2$ , dato che il nostro gemello ovviamente non si muove rispetto a se stesso, cosicché per lui $\Delta x_0^2 = 0$, e quindi $\Delta s_0^2 = c^2 \Delta t_0^2$. Ma i due $\Delta s^2$ devono essere uguali: $\Delta s_1^2 = \Delta s_0^2$. Ne segue subito che $c^2 \Delta t_1^2 - v^2 \Delta t_1^2 = c^2 \Delta t_0^2$, ovvero che $\Delta t_1^2 (c^2 - v^2) = c^2 \Delta t_0^2$, e quindi

$$\Delta t_1^2 (1 - v^2/c^2) = \Delta t_0^2 .$$

Si vede subito che il tempo $\Delta t_0$ trascorso per il nostro gemello (tra i due eventi considerati della sua vita) è *minore* del tempo misurato per noi, essendo uguale al nostro $\Delta t_1$ moltiplicato per un fattore [la radice quadrata di $1-v^2/c^2$] chiaramente minore di 1; in altre parole, usando per una volta i differenziali invece delle quantità finite, si potrà scrivere:

$$dt_o = dt \sqrt{1 - \frac{v^2}{c^2}}$$

Viceversa, possiamo scrivere che il tempo $\Delta t_1$ misurato da noi sarà uguale al tempo $\Delta t_0$ misurato da lui *diviso* per la la radice quadrata di $1-v^2/c^2$: si trova sempre che, secondo noi, il nostro gemello (durante quel tragitto) ha speso un tempo minore di quello passato per noi; e così è rimasto più giovane.

Questo fatto è *relativo*, nel senso che il gemello [considerando se stesso fermo e *noi* in moto!] valuterà, e calcolerà, esattamente il contrario di noi: e riterrà che siamo noi a restare più giovani. Ma ad ogni osservatore tale fatto (la "dilatazione del tempo" per osservatori od oggetti in moto) sarà necessario per descrivere correttamente il mondo fisico da lui osservato. E infatti tale **"dilatazione del tempo"** è stata verificata sperimentalmente un grandissimo numero di volte. In particolare, nel caso dei "muoni" prodotti dai raggi cosmici nell'alta atmosfera e dotati di velocità $v = 0.99\ c$ (sicché la radice quadrata di $1-v^2/c^2$ vale 00.1, cioè 1/100), si è verificato che essi appaiono a noi possedere una vita *cento volte* maggiore di quella che presentano quando sono fermi (e che tali muoni effettivamente vivono, essendo fermi rispetto a se stessi)...

Quanto sopra potrebbe apparire paradossale. Ma ciò non è, dato che si può dare una risposta alla domanda "ma chi resta davvero più giovane?". Occorre però che i due gemelli *si incontrino una seconda volta*, così da confrontare oggettivamente la loro età.

Per esempio, se il gemello sul razzo inverte il suo moto (per semplicità, mantenendo in modulo sempre la medesima velocità) e ci incontra in un secondo evento P, la relazione differenziale sopra scritta si integra banalmente [essendo essa lineare tra $dt_0$ e $dt_1$, e rimanendo costante la radice, che dipende dal quadrato (e non dal segno) di $v$ ], per cui la relazione $\Delta t_1^2 (1- v^2/c^2) = \Delta t_0^2$ ora ha un significato assoluto: $\Delta t_0$ e $\Delta t_1$ rappresentando adesso il tempo trascorso, tra il primo e il secondo incontro, rispettivamente per il gemello sul razzo e noi sulla terra. Ovvero, il gemello sul razzo sarà oggettivamente più giovane di noi. Questa volta, però, non si è di fronte a risultati paradossali, dato che i due gemelli si sono trovati in situazioni fisiche *differenti*: il gemello sul razzo ha dovuto abbandonare il suo sistema di riferimento inerziale, accelerando per tornare indietro; mentre quello vissuto sulla Terra non ha sostanzialmente abbandonato il suo sistema inerziale. [L'esame della Figura 4 evidenzia come il cammino rettilineo OA (lungo l'asse *t*) implica la maggiore durata temporale. Questo è dovuto tra parentesi, al fatto che la geometria dello spazio-tempo (a causa di quel segno *meno*) non è euclidea, ma –come si dice—pseudo-euclidea].

Ovviamente, se il gemello sul razzo non torna indietro, ma siamo noi ad accelerare per raggiungerlo, tra i due eventi-incontro O e B il tempo più lungo sarà passato per lui!; e noi saremo i più giovani (anche di un fattore cento, come si disse, se la nostra velocità sarà di 0.99 volte quella *c* della luce): ved. Figura 5.

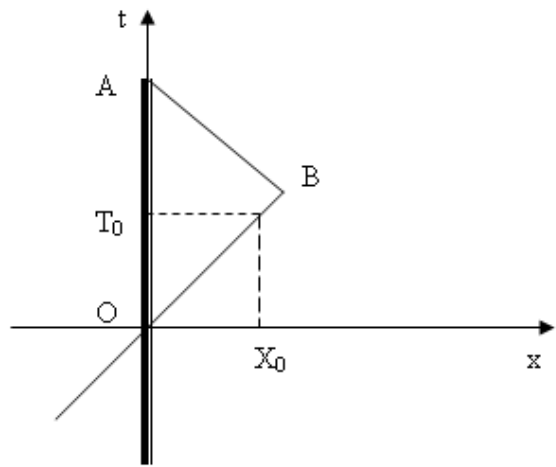

*Fig. 4*

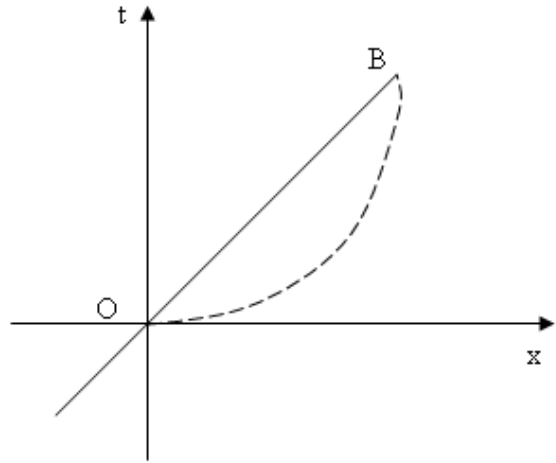

*Fig. 5*

*Le Categorie di Spazio e Tempo, e Spazio-Tempo:*

Einstein, in altre parole, mise in evidenza che i nostri innati concetti di spazio e tempo dipendono dalla natura dell'ambiente in cui è avvenuta l'evoluzione biologica su questa Terra, e dalla limitatezza delle esperienze che tale ambiente ha permesso. Detta evoluzione ha fatto sì che il nostro cervello inserisca le sorgenti delle nostre sensazioni all'interno di uno spazio piatto tridimensionale, con un tempo unidimensionale; essendo questa la teoria che è risultata la più semplice ed efficace per interpretare la realtà intorno a noi. [Ricordiamo che lo stesso fenomeno della nostra visione è una pura creazione del nostro cervello, dato che all'esterno di noi esistono solo oggetti che inviano delle onde elettromagnetiche alla nostra retina: onde costituite, appunto, da campi elettrici e campi magnetici che non sono di per sé né luminosi né colorati]. Se i nostri pronipoti avranno esperienza di velocità vicine a quella della luce, nasceranno forse col concetto innato di spazio-tempo. E, se vivranno vicino a grosse concentrazioni di masse (l'intera Terra ha una massa che potrebbe stare in una sfera di 130 metri di diametro, se avesse la densità della materia nucleare), nasceranno probabilmente con l'innata capacità di intuire e rappresentarsi lo spazio-tempo incurvato dalla presenza di materia. Tutto ciò può essere *confrontato con gli* "elementi puri a priori" di Kant.

## Come Einstein arrivò a tale risultato? L'ha narrato lui stesso.

Giunto a conoscenza degli esperimenti che mostravano la velocità *c* della luce nel vuoto essere *la medesima* per tutti gli osservatori inerziali, Einstein si era chiesto invano per quasi un anno come ciò potesse accadere, contro le leggi della fisica classica:

*"Per caso mi aiutò a venirne fuori il mio amico italiano di Berna, Michele Besso. Era una bellissima giornata quando gli feci visita con questo problema in mente. Cominciai la conversazione dicendogli subito: 'Di questi tempi ho lavorato su di un problema difficile. Oggi sono venuto qui per combattere contro di esso con te".*

Discutemmo insieme ogni aspetto del problema. Improvvisamente io compresi quale ne fosse la chiave. Il giorno seguente tornai e, senza neppure salutarlo, gli dissi:

*"Grazie. Ho completamente risolto il problema. La mia soluzione era stata un'analisi del concetto di tempo..."*

La conseguenza più clamorosa di questa nuova analisi del concetto di tempo, tra quelle note, è stata messa in evidenza, più sopra, attraverso il cosiddetto *paradosso dei due gemelli*. Come già si è detto, questo effetto, per quanto insolito, è stato verificato innumerevoli volte, anche inviando degli orologi a viaggiare su un aereo e confrontando alla fine del viaggio questi orologi con degli orologi identici rimasti a terra. Gli orologi viaggiatori --cioè non inerziali-- sono risultati più giovani (segnavano un tempo inferiore), e nella misura prevista da Einstein, degli orologi sedentari. Possiamo quindi essere sicuri che, il giorno in cui faremo un viaggio di andata e ritorno fino alla stella più vicina con una velocità, diciamo, pari a tre quarti di quella della luce, al ritorno troveremo i nostri parenti e amici invecchiati oltre 4 anni più di noi: per i "terrestri" saranno infatti passati 12 anni, e per noi neanche 8 anni... Ripetiamo che la nostra vita, della durata, diciamo, di cento anni, apparirà come di centocinquanta anni a un osservatore che si muova rispetto a noi con la velocità di 200 mila km/s. Così la nostra massa, diciamo di 70 chili, potrà essere ritenuta di oltre 100 chili da un altro osservatore... Questi aspetti *relativi* della Fisica messi in luce da Einstein, non possono non richiamare molti aspetti dell'opera pirandelliana, da "*Il fu Mattia Pascal*" a "*Uno, nessuno, centomila*" (anche se poi Einstein ci insegnò a ricostruire la fisica sulla base di quantità assolute o invarianti).

## *La Relatività Generale:*

Nel recente 2005 si sono celebrate le straordinarie pubblicazioni einsteiniane del 1905, e in seguito riparleremo delle più famose: quelle sulla Relatività Speciale. Ma se ci è permesso un salto in avanti, vorremmo accennare alla Relatività Generale (RG) e ai suoi sviluppi più recenti, perché con la RG l'orizzonte si amplia ancora di più. Premettiamo che la vecchia fisica dava per scontato che la massa inerziale m, che entra nella legge fondamentale della dinamica, $F=ma$, coincidesse con la massa gravitazionale (che entra invece nella legge della gravitazione universale, $F=GmM/r^2$): è proprio questa coincidenza che fa sì che tutti i corpi sulla Terra cadano con la medesima accelerazione $g$ di gravità, indipendentemente dalla loro massa. Partendo da questa semplice considerazione, Einstein comprese che la gravitazione può interpretarsi come dovuta non tanto ad un "campo gravitazionale" sovrapposto ad uno spazio-tempo piatto e infinito, quanto ad una *deformazione* dello spazio-tempo stesso, che viene incurvato dalla presenza di masse. La Terra, ad esempio, si muoverà intorno al Sole semplicemente perché descriverà una "retta" (o meglio una geodetica) in uno spazio-tempo leggermente incurvato dalla massa del Sole [si noti come sia essenziale qui considerare l'incurvamento del tempo, e non solo dello spazio]. Così Einstein spiegò vari fenomeni astronomici e cosmologici fino ad allora privi di interpretazione (a volte basandosi, semplicemente, su uno dei suoi famosi "*esperimenti concettuali*").

A seguito di detta curvatura (dello spazio, ad esempio), l'intero cosmo può diventare curvo, e magari chiuso su se stesso e quindi finito. Anche tali conseguenze possono essere messe a confronto con la cosmogonia e cosmologia kantiana.

Einstein stesso raccontò come gli nacque la prima idea della Relatività Generale (RG). Stava meditando sulla connessione tra inerzia e peso:

*"Sentivo che questo problema non poteva essere risolto all'interno della Relatività Speciale. L'idea buona mi venne un giorno improvvisamente. Sedevo su una sedia del mio Ufficio Brevetti di Berna. Improvvisamente un pensiero mi colpì: se una persona cade in caduta libera essa non sentirà il proprio peso... Questo semplice esperimento concettuale esercitò su di me una profonda impressione. Mi portò alla teoria della gravità. Io continuai il mio pensiero: un uomo che cade si muove di moto accelerato; quindi ciò che lui vede e giudica vale in un sistema di riferimento accelerato. Decisi di estendere la Relatività Speciale ai riferimenti accelerati. Sentivo che così facendo avrei automaticamente risolto il problema della gravità. Se un uomo che cade non sente più il proprio peso, ciò significa che nel suo sistema di riferimento nasce un nuovo campo gravitazionale che cancella quello dovuto alla Terra. Nel riferimento accelerato deve nascere dunque un nuovo campo gravitazionaIe. Non riuscii allora a risolvere il problema completamente. Mi ci vollero infatti altri otto anni per ottenere finalmente la soluzione completa..."*

Una delle soluzioni possibili della RG è che il nostro cosmo sia soggetto a cicli di espansione e contrazione, ponendo su un piano scientifico la teoria del "big bang ciclico", che in precedenza era solo una ipotesi, già elegantemente espressa verso l'anno 100 a.C. (!) dall'antico scienziato greco Posidonio:

*"L'universo è dominato da una forza immensa... e, seguendo le trasmutazioni fisiche, ora si contrae consumato dal fuoco, ora si espande dando nuovamente inizio alla creazione del mondo"*.

Ripetiamo che, però, è solo con Einstein che il problema cosmologico è divenuto, per la prima volta nella storia umana, una questione non più di natura soltanto speculativa e filosofica, ma all'altezza delle capacità scientifiche dell'uomo.

Aggiungiamo che Einstein, come si è visto, ha mostrato che il più adatto sfondo onde collocarvi gli oggetti (sorgenti delle nostre sensazioni) è lo spazio-tempo, in cui spazio e tempo sono tra loro legati, e non più indipendenti. E' solo nello spazio tempo, infatti, che le distanze tra due eventi sono universali, cioè uguali per tutti gli osservatori.

Quanto sopra contrasta con quanto teorizzato da Kant? In un certo senso, no: perchè noi esseri umani, formati dall'interazione col mondo accessibile in passato alla nostra esperienza (nel quale le velocità in gioco sono trascurabili rispetto a quella della luce), continuiamo a usare come elementi a priori e innati quelli di spazio e tempo (e solo successivamente, con una ragione "non immediata" e "non intuitiva", passiamo alle distanze invarianti, spazio-temporali); a ciò siamo obbligati dalla stessa natura dei nostri sensi, e dei nostri attuali strumenti scientifici. Possiamo prevedere, però, che i nostri pronipoti –-se posti continuamente a contatto con velocità prossime a quelle della luce— potranno modificare le proprie categorie innate, e nascere con l'intuizione immediata del nuovo elemento a priori "spazio-tempo". Kant accetterebbe una previsione di questo tipo? Pensiamo di sì (anche se la sua filosofia non poteva ovviamente contenerla), dato che Kant ha sempre dato molta importanza alla evoluzione progressiva, anche dei giudizi sintetici a priori (in ciò pure distaccandosi nettamente dai Razionalisti).

Parlando della Relatività Generale abbiamo detto che essa ha permesso di affrontare scientificamente il problema della forma e nascita del nostro Cosmo. Ma Kant, limitando ovviamente la sua immagine di Cosmo a quanto allora conosciuto, ovvero al Sistema Solare, aveva contribuito notevolmente ai tentativi di spiegare la formazione del Sole e dei pianeti; infatti, già verso la metà del '700, Kant –ispirandosi alle idee di Newton— ipotizza che il sistema solare si sia formato a partire da una nebulosa primordiale, dalla quale Sole e pianeti sono nati per condensazione gravitazionale di polveri e gas. Qualche anno più tardi, e in maniera indipendente, il fisico-matematico francese Pierre

Simon de Laplace ha una simile idea, dando alla teoria di Kant-Laplace solide basi fisiche. Tra parentesi, la nube primordiale di Kant era fredda e immobile; secondo Laplace, invece, essa era calda e in rotazione. Nella formulazione laplaciana della teoria kantiana, la contrazione della nebulosa, accompagnata dal conseguente noto aumento della velocità di rotazione, dà vita ad un nucleo (da cui si forma il Sole) e ad un disco, prodotto dall'espulsione dalla regione equatoriale del nucleo di materia spinta verso l'esterno dalla cosiddetta forza centrifuga. Il disco è formato da una serie di anelli simili a quelli di Saturno, i quali si frammentano in parti che successivamente, aggregandosi, producono i pianeti.

In ogni caso, per quanto riguarda i pianeti, l'idea originale di Kant e di Laplace è che essi fossero nati a partire da irregolarità presenti nell'atmosfera del Sole, che doveva estendersi molto lontano nello spazio. Mentre Laplace propendeva per una nube di gas, Kant parlava di una nube di generiche particelle; ma entrambi pensavano, ripetiamolo, che il materiale residuo alla formazione del Sole si fosse suddiviso in a una serie di anelli concentrici, situati a distanze via via crescenti dall'astro centrale; e che dal materiale presente in tali anelli si sarebbero poi condensati i diversi pianeti.

Vale ancora la teoria di Kant e Laplace? Nonostante esistano ora varie teorie sulla nascita del sistema solare, sostanzialmente le sue idee informatrici sembrano rimanere valide. Anzi, teorie del genere hanno un ruolo scientifico, ora, per la spiegazione dell'origine delle galassie, per collasso gravitazionale delle nubi di Idrogeno (ed Elio) che formavano l'universo primordiale.

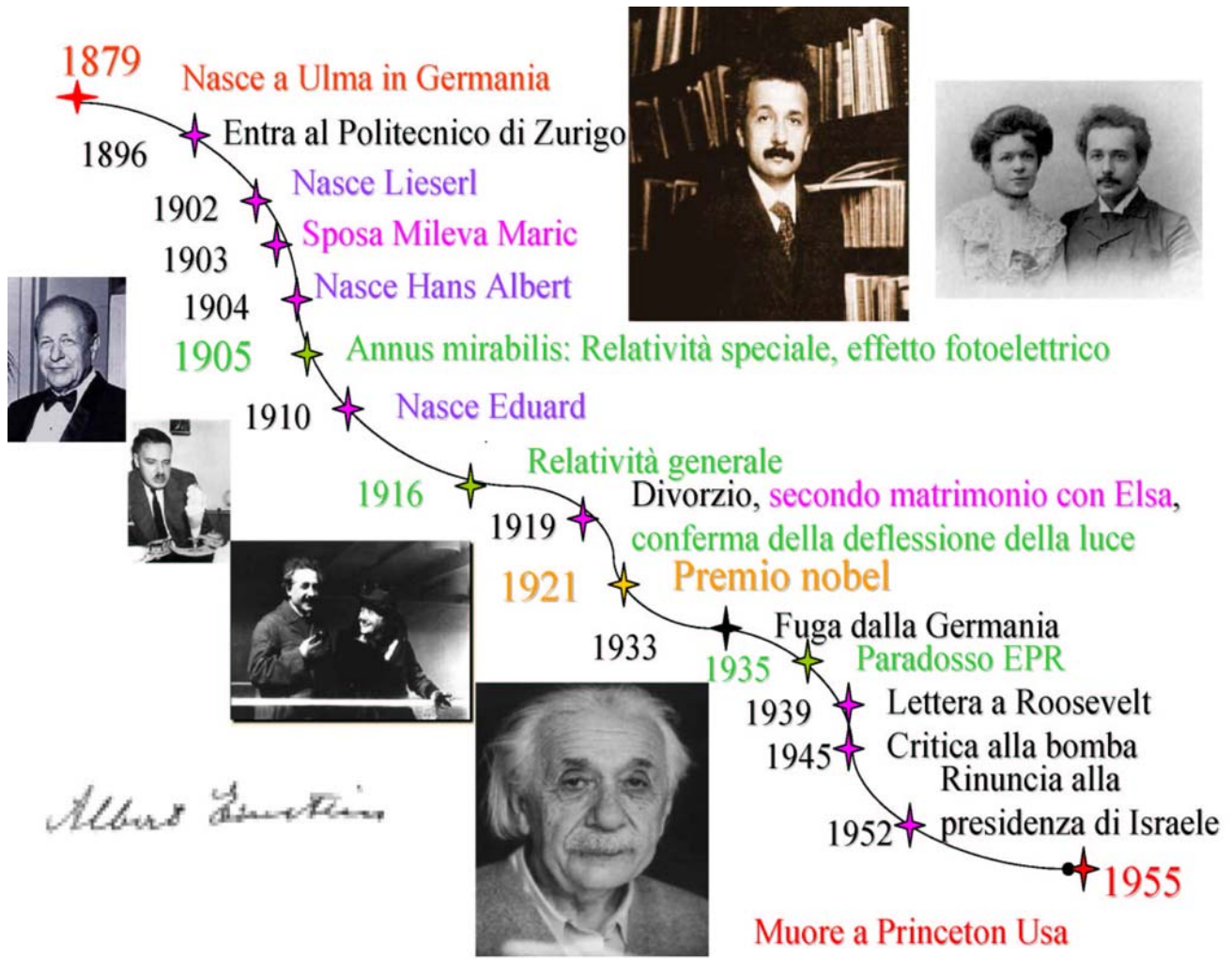

*Fig.6*

*La proporzionalità tra massa totale ed energia totale di un corpo (E=mc²).*

Nel 1905, sulla Relatività Speciale, Einstein scrisse **due** articoli: nel secondo –di un paio di pagine-- derivò la sua nota equazione *E=mc²*, la quale afferma il fatto interessante che *l'inerzia* di un corpo è proporzionale al suo contenuto di *energia*; o, in altre parole, che la massa totale di un corpo è proporzionale al suo contenuto in energia, *e viceversa*. [In particolare, un oggetto fermo (e privo, oltre che di energia cinetica, anche di energia potenziale), avrà un contenuto di energia $E_0 = m_0 c^2$, ove $m_0$ è la cosiddetta massa-a-riposo dell'oggetto: ovvero una energia elevatissima, dato l'alto valore di *c*]. La suddetta relazione, pur essendo teorica, è alla base: 1) dei processi di fissione di nuclei atomici pesanti, che hanno portato, a Chicago, nel 1942, ad una epocale applicazione pacifica (la "pila atomica") operata da Enrico Fermi, e successivamente, purtroppo, a terrificanti applicazioni belliche (la bomba A, o "bomba atomica"); e 2) dei processi di fusione di nuclei leggeri, che la natura ha scelto per fare produrre energia alle stelle (e al Sole), ma che l'umanità non ha ancora saputo o voluto controllare (mentre ne ha prodotto solo una versione bellica, spaventosa: la bomba H, o "bomba all'Idrogeno").

Succede infatti che, *fondendo* due nuclei di atomi *leggeri* (fino a formare un nucleo di Ferro, e non oltre), il prodotto ha una massa a riposo *minore* delle due masse a riposo di partenza; così che si rende disponibile una certa energia, di solito emessa mediante la generazione di raggi gamma $\gamma$ (onde elettromagnetiche, anzi, fotoni di altissima frequenza). Al di là dell'elemento Ferro, però, avviene il contrario: è solo *scindendo* un nucleo pesante (ad esempio, di Uranio) che si ottengono prodotti con massa a riposo la cui somma è minore di quella di partenza; quindi, nel caso di nuclei *pesanti,* è solo con la *fissione* che si possono produrre raggi $\gamma$ [anzi, nel caso della fissione del nucleo dell'Uranio vengono prodotti anche neutroni che colpiscono e spaccano altri nuclei, potendo dare avvio a una reazione a catena].

La natura usa la fusione nucleare per far sì che le stelle, a partire dall'Idrogeno primordiale, fabbrichino nella loro regione centrale i nuclei degli elementi successivi, nel contempo emettendo energia $\gamma$. Vediamo cosa avviene nel nostro Sole (anche questa parentesi può essere, volendolo, "saltata")

I PROCESSI DI FUSIONE NUCLEARE CHE AVVENGONO
 AL CENTRO DEL SOLE

Nell'economia del Cosmo, il Sole –come tante altre stelle— ha attualmente il semplice compito di trasformare nuclei dell'elemento numero 1, l'Idrogeno (che sono protoni, $p^+$, che si potrebbero indicare con $H^+$) in nuclei dell'elemento numero 2, l'Elio (che sono le cosiddette particelle alfa, che possiamo indicare con $He^{++}$, costituite da due protoni $p^+$ e due neutroni $n^0$). I necessari processi di fusione occorrono però solo a temperature di milioni di gradi. Infatti i protoni, per esempio, si respingono, dato che entrambi sono carichi positivamente; ed occorrono velocità grandissime affinché si avvicinino a distanze minori di 1 fermi (1 fm = $10^{-13}$ cm), quando entrano in gioco le forze attrattive nucleari, molto più potenti della repulsione elettrica. Ma velocità grandissime vogliono dire temperature enormi. Al centro del Sole la temperatura è di circa 15 milioni di gradi, sufficienti perché avvenga in abbondanza la seguente serie di reazioni nucleari esotermiche (e quindi spontanee, con produzione di energia, così che si parla di reazioni termonucleari):

a) $p^+ + p^+ \rightarrow d^+ + e^+ + \nu$

ove $d^+$ rappresenta il deutone (nucleo dell'Idrogeno pesante: costituito da un protone e un neutrone); in questa reazione nucleare un protone si è trasformato in neutrone, producendo un elettrone positivo ---o antielettrone--- $+ e^+$, che si porta via la carica elettrica, e un neutrino elettronico $\nu$: secondo la reazione $p^+ \rightarrow n^0 + e^+ + \nu$;

b) $d^+ + p^+ \rightarrow He_3^{++} + \gamma$
ove il deutone e il protone si fondono, producendo il nucleo di $He_3$, ed emettendo un raggio gamma

c) a questo punto due nuclei di $Elio_3$ si fondono in modo da creare un nucleo di normale $Elio_4$, riproducendo due protoni:
$He_3^{++} + He_3^{++} \rightarrow He_4^{++} + 2\, p^+$.

In totale, si sono consumati quattro protoni (nuclei di Idrogeno), costruendo un nucleo di Elio, con produzione di $e^+$, $\nu$, e $\gamma$. Gli elettroni positivi (anti-elettroni, o positroni) vengono subito assorbiti dal materiale del Sole; i neutrini se ne vanno attraversando il Sole senza alcuna interazione (per fermarne la metà, occorrerebbe un muro di piombo spesso da qui alla stella più vicina a noi, dopo il Sole: cioè, spesso circa 4,5 anni-luce); quello che a noi interessa è la produzione di energia radiante, ovvero dei $\gamma$. Per fortuna, i gamma impiegano decine di migliaia di anni prima di arrivare alla superficie del Sole, perdendo energia e trasformandosi via via nelle particelle di luce emesse dalla superficie del Sole (la quale corrisponde solo a circa 6 mila gradi, e non più 15 milioni di gradi!). Il Sole emette infatti lo spettro caratteristico (col massimo in corrispondenza del colore giallo) di un corpo a 6 mila gradi; e i nostri occhi, tra parentesi, si sono adattati a divenire sensibili alle radiazioni emesse dal Sole in maggior copia (che sono quelle che vanno dal rosso al violetto).

Nelle stelle più calde del Sole, che abbiano al loro centro temperature maggiori di 20 milioni di gradi, il processo più abbondante per la produzione di Elio a partire dall'Idrogeno non è quello visto sopra (che domina nel Sole), ma il cosiddetto "ciclo di Bethe". Tale processo è più complicato, ma, a partire da 4 nuclei di Idrogeno, produce ancora un nucleo di Elio, usando come *catalizzatori* nuclei di Carbonio (i quali non si consumano, ma si riformano al termine del ciclo). Come mai tali stelle contengono Carbonio, se esse stanno semplicemente consumando Idrogeno per produrre Elio? I nuclei di tutti gli elementi vengono forgiati solo all'interno di quelle fucine nucleari che sono i nuclei stellari! (a parte i nuclei degli elementi più pesanti del Ferro, i quali si formano probabilmente durante le fasi di esplosione di supernova).. Cio vuol dire che le stelle suddette non sono stelle di prima generazione, ma sono nate dalla condensazione di nubi di Idrogeno arricchite dai prodotti di stelle precedenti... A un certo punto della vita di ogni stella (che dipende ad esempio dalla sua massa), essa, quando termina il combustibile nucleare usato in quella sua fase, esplode come Super-nova disperdendo nello spazio i nuclei da lei fucinati al proprio interno. Il nostro Sole, quando terminerà l'Idrogeno, non esploderà (lo farà in fasi successive), perché la sua massa è abbastanza piccola da permettergli di non divenire instabile. Si calcola ad ogni modo che il nostro Sole stia trasformando Idrogeno ed Elio da 5 miliardi di anni, e continuerà per altri 5 miliardi di anni.

Abbiamo parlato di "interno" del Sole. Spieghiamoci meglio. L'interno del Sole è formato da un nucleo, col raggio di circa 150 mila km, nel quale le temperature sono tali (circa 15 milioni di gradi, come abbiamo detto) da permettere la fusione termonucleare, e la generazione di raggi gamma. La radiazione gamma, per fortuna, viene assorbita e riemessa innumerevoli volte nella "zona radiativa" che avvolge il nucleo, e che si estende per ben 450 mila km. Successivamente la distribuzione dell'energia avviene per movimento delle masse solari: si ha, cioè, una zona convettiva, che copre però solo gli ultimi 10 mila km prima della superficie solare. La superficie visibile del Sole si chiama *fotosfera*: essa, che è la parte sommitale della zona convettiva, è l'involucro che irradia la luce solare; trovandosi a circa seimila gradi Kelvin, irraggia soprattutto quelle onde elettromagnetiche che per noi sono diventate luce visibile ai nostri occhi (con un massimo di intensità corrispondente al colore giallo). Durante le eclissi di Sole, diventa visibile anche l'atmosfera solare, in particolare la cromosfera La parte più esterna dell'atmosfera solare, la corona, diventa invece visibile solo durante le eclissi totali (ma appare estendersi fino a 17 *milioni* di km dal Sole.

La massa del Sole è circa 300 mila volte quella della Terra, e il suo diametro è circa 109 volte quello terrestre. La sua potenza di irradiazione è di circa $3,8 \times 10^{23}$ kW. Ogni secondo si fondono, per produrre –come abbiamo visto sopra— elio, circa 565 milioni di tonnellate di idrogeno; una percentuale pari allo 0,7% di tale materia si trasforma in energia raggiante, così che la massa che resta nel Sole diminuisce ogni secondo di circa 4,5 milioni di tonnellate.

# PARTE SECONDA

## ALTRI COMMENTI SU ALCUNE CONSEGUENZE PEDAGOGICHE (E "FILOSOFICHE") DELL'OPERA DI A.E.

*Altre conseguenze pedagogiche (e"filosofiche") dell'opera einsteiniana:*
La pubblicazione da parte del ventiseienne Albert Einstein, durante il solo anno 1905, dei suoi mirabili quattro articoli –e qui ci riferiremo soprattutto ai due suoi lavori che hanno dato avvio alle Teorie Relativistiche— ha operato profondi mutamenti sulla nostra visione del mondo che ci circondo. Tanto che ora sono alla portata scientifica dell' umanità quesiti come i seguenti:

Il tempo è unidimensionale? Sono proprio tre le dimensioni dello spazio? Quante dimensioni spaziali e temporali può avere il mondo in cui viviamo? E il nostro spazio, per la presenza di materia, si deforma acquistando una curvatura? Sono alcune delle domande a cui parte della ricerca contemporanea sta tentando di rispondere, pur sotto il velo del formalismo fisico-matematico.

Ma Einstein ha rivoluzionato ancor più il nostro modo di fare e vedere la scienza, e il mondo, dal punto di vista *"filosofico"*. Prendiamo quindi il discorso più alla lontana...

Premettiamo che l'uomo da quando esiste (ovvero da molti milioni di anni, come suggeriscono le scoperte africane, ad esempio in Tanzania e in Etiopia) ha sempre agito sulla natura, e osservato la realtà, ricorrendo agli strumenti sensoriali e logici ereditati da almeno tre miliardi di anni di evoluzione biologica precedente. Il maggior risultato di questo continuo sforzo è stato la creazione del linguaggio: creazione in cui è facile riconoscere il contributo di tutte le facoltà umane, e tra i primi il contributo dell'atteggiamento scientifico (si pensi al carattere inter-soggettivo del linguaggio). In un certo senso la scienza moderna non fa altro che ampliare (con telescopi e microscopi) e approfondire (con misurazioni precise) il campo della nostra esperienza, per costruire quindi un supplemento di linguaggio atto a descrivere (e "capire") i nuovi mondi di esperienza. Da questo punto di vista l'attività scientifica dell'uomo presenta nei millenni forti caratteri di continuità. Essa, inoltre, è in una certa misura spontanea e universale: già da poppanti di fronte al fenomeno della gravità cominciamo tutti a fare prove e riprove, utili e necessarie, sul modo di agire di questa forza...

La scienza non è, poi, nettamente separabile dalle altre attività umanistiche: essa è una creazione dello spirito umano; Einstein, anzi, parla di potremmo una libera creazione: infatti, nel 1933 dichiarò:

*"I postulati e i concetti fondamentali sui quali si basa la fisica teorica sono libere invenzioni dell'intelletto..., e costituiscono la parte essenziale di una teoria, la parte che la logica non può toccare".*

La scienza, infatti, viene sviluppandosi con caratteristiche proprie, ma richiede pure doti di creatività, intuizione, fantasia, senso estetico. La scienza, dunque, al pari degli altri prodotti dell'uomo si è continuamente sviluppata durante l'evoluzione umana. Dal diciassettesimo secolo (anche per la diffusione della stampa), la scienza ha assunto, poi, con evidenza il proprio carattere di essere fondata sulla cooperazione di molti: ciò era stato già ben compreso da Seneca, il quale --dopo aver discusso delle comete-- concluse con le nobili parole:

*"Accontentiamoci di ciò che abbiamo finora scoperto. Quelli che verranno dopo di noi aggiungeranno il loro contributo alla verità".*

Anche Einstein non considerò le proprie equazioni della Relatività Generale come l'ultima parola, chiamandole anzi effimere, e in una lettera scrisse:

*"Tu immagini che io guardi indietro alla mia vita con calma soddisfazione. Ma non c'è un singolo concetto della cui incrollabilità io sia convinto, e non so se in generale io sono sulla strada giusta".*

Più precisamente, è ben noto che il pensiero scientifico procede per successive approssimazioni, e che la teoria nuova contiene la vecchia come caso particolare. La Relatività Speciale, così, si riduce alla precedente Meccanica di Galilei-Newton quando le velocità in gioco sono molto più piccole della velocità della luce. E la Relatività Speciale lascerà certamente il posto a una teoria più ampia, che la conterrà come caso particolare.

E' stato comunque l'esempio portoci da Einstein, attraverso la sua opera e vita scientifica, quello che più ha influenzato nei fini e nei mezzi l'intera fisica del secolo appena trascorso, così come continuerà certamente a fare nel secolo presente.

*L'esempio di Einstein*
La grande ventata di freschezza apportata da Albert Einstein ha riguardato anzitutto le sorgenti e le caratteristiche della grande scienza; ricordiamo dunque alcuni caratteri delle vie che furono aperte da Einstein, senza dimenticare che il suo insegnamento (e l'esempio stesso della sua vita di pensiero) è *anche* prezioso dal punto di vista della *pedagogia* e della educazione.

a) Einstein ci ha ricordato che fare scienza vuoi dire non certo redigere un catalogo di fatti (anche se questo può essere un necessario passo preliminare), bensì "comprenderli", scoprendo ciò che resta costante nel loro manifestarsi e divenire. Einstein, anzi, un giorno osservò che non poteva capire

*"come mai qualcuno potesse sapere così tante cose, e capire così poco"*,

e sempre sottolineò il pericolo di conoscere troppi fatti e di perdervisi in mezzo. Invece la scienza tende a collocare i fenomeni in strutture ordinate, così da poterli descrivere in una maniera estremamente più elegante, logica e compatta che non per elenco. A tale scopo non aiutano molto le deboli forze dell' induzione. Einstein dichiarava nelle sue *Note autobiografiche:*

*"Una teoria può essere verificata dall'esperienza, ma non esiste alcun modo per risalire dall'esperienza alla costruzione di una teoria. Equazioni di tale*

*complessità come sono le equazioni del campo gravitazionale possono essere trovate solo attraverso la scoperta di una condizione matematica logicamente semplice, che determini completamente, o quasi completamente, le equazioni. Una volta in possesso di condizioni formali abbastanza stringenti, non c'è bisogno di una grande conoscenza dei fatti per costruire una teoria".*

Aggiungiamo che Einstein ricorse spesso addirittura ad esperimenti puramente concettuali *(gedankenexperimente)*;

b) Einstein ci ha ricordato col suo stesso esempio che la scienza per rinnovarsi ha bisogno della freschezza delle idee dei giovani. La scuola tende a volte a squadrare troppo la mente dei giovani in una forma preconfezionata. Einstein ebbe la fortuna di possedere fortissimo l'istinto di proteggere l'originalità della sua mente. In seconda ginnasio, anzi, troncò gli studi scolastici per un anno raggiungendo la famiglia a Milano, e vagabondando ad esempio --sembra a piedi-- fino a Genova, ove aveva dei parenti. Riprese poi gli studi, ma in Svizzera. Dei quattro anni di studio all'ottimo Politecnico di Zurigo ha lasciato scritto:

*"Per superare gli esami, volenti o nolenti, bisognava imbottirsi la mente di tutte queste nozioni... Una simile coercizione ebbe su di me un effetto così scoraggiante che, dopo aver superato l'esame finale, **per un anno intero mi ripugnò prendere in considerazione qualsiasi problema scientifico...**"*

Altrove dirà, anche se con riferimento alla musica:

*"Io ritengo, tutto sommato, che l'amore sia un maestro più efficace del senso del dovere"*;

c) Einstein ci ha pure ricordato che uno dei motori dello scienziato è la fede in una *unità razionale* del mondo, unità che almeno in parte è percepibile dalla nostra mente. Tra parentesi, la scienza cosiddetta moderna è nata, o rinata, in Europa forse perché qui il terreno vi fu in parte preparato (oltre che dalle importanti attività artigianali) dagli studi medievali ispirati alla concezione di un Dio *unico*. Infatti, per giudicare una teoria, Einstein spesso si domandava se --qualora lui fosse stato Dio-- avrebbe creato il Cosmo in quel modo. È ben noto ad esempio come Einstein (nel rifiutare la visione probabilistico-quantistica dei fenomeni microfisici) dichiarasse più volte di non credere che "*Dio giocasse ai dadi*";

d) Abbiamo già visto quanto rilievo abbia dato Einstein al processo intuitivo rispetto a quello induttivo. Ritorniamo su questo aspetto dell'insegnamento einsteiniano riportando alcune delle parole da lui pronunciate nel lontano 1918:

*"Compito supremo dello scienziato è di pervenire a quelle leggi elementari universali partendo dalle quali il cosmo può essere costruito con la pura deduzione. Non esiste alcun sentiero logico che conduca a queste leggi: soltanto l'intuizione, appoggiata ad una sensata comprensione della realtà, può condurre ad esse... L'anelito a contemplare l'armonia cosmica è la fonte della pazienza inesauribile e della perseveranza del vero uomo di scienza... Lo sforzo quotidiano scaturisce non già da un'intenzione deliberata o da un programma, ma direttamente dal cuore"*.

Prima di passare alla Parte Terza, ove si affronta più in dettaglio il problema della responsabilità degli intellettuali, ricordiamo ancora una volta i rapporti di Seneca col potere del suo tempo.

# PARTE TERZA

## SCIASCIA, MAJORANA, AMALDI, EINSTEIN E LA RESPONSABILITA' DEGLI INTELLETTUALI

***L'intervento di Einstein presso il Presidente degli USA, Roosvelt:***
Albert Einstein, pur essendo uomo pacifico, ebbe però un ruolo importante per convincere gli Americani a costruire la Bomba atomica (Bomba A). Era da anni molto amico di Leo Slizard, un ungherese che (come Einstein) aveva abbandonato la Germania di Hitler. Agli inizi del 1939, si seppe che si poteva produrre una reazione a catena nell'Uranio, scindendone un numero rapidissimamente crescente di nuclei. La "fissione" (scissione) di ogni nucleo produceva una piccola energia –comunque dell'ordine di 1 MeV, cioè un milione di volte maggiore di quella prodotta da un atomo in una reazione chimica--, ma produceva *due* neutroni, ciascuno dei quali poteva scindere un altro nucleo, così che in breve l'energia liberata in totale poteva divenire enorme. Il 2 di agosto del 1939 Slizard convince Einstein a scrivere una lettera [riprodotta qui di seguito] al Presidente degli Stati Uniti, Roosvelt, informandolo del pericolo che la reazione a catena venisse usata da Hitler allo scopo di produrre un'arma un milione di volte più potente di quelle convenzionali. Roosvelt lasciò quasi cadere la proposta. Contemporaneamente, l'italiano Enrico Fermi continuava negli Stati Uniti le properie ricerche romane, giungendo, fortunatamente, novello Prometeo, ad insegnare all'umanità l'uso *pacifico* del "fuoco nucleare" (Chicago, 1942). Un giorno prima dell'attacco giapponese a Pearl Harbor, col quale il Giappone entrò in guerra con gli USA, Roosvelt aveva però alfine deciso di fare partire a Los Alamos (New Mexico) il mastodontico "Progetto Manhattam" per la costruzione della Bomba A.

# TRADUZIONE DELLE LETTERA DI EINSTEIN AL PRESIDENTE ROOSVELT:

**Albert Einstein**
**Old Grave Road**
**Nassau Point**
**Peconic, Long Island**

**2 agosto 1939**

**F.D.Roosvelt**
**Presidente degli Stati Uniti**
**Casa Bianca**
**Washington, DC**

Sir:

Recenti risultati di E.Fermi e L.Szilard, che mi sono stati trasmessi in forma manoscritta, mi inducono a prevedere che l'elemento uranio possa essere convertito nel prossimo futuro in una potente sorgente di energia. Alcuni aspetti della situazione che di conseguenza si è aperta paiono richiedere attenzione e, se necessario, una rapida azione da parte del Suo Esecutivo. Considero pertanto mio dovere portare alla Sua attenzione i seguenti fatti e raccomandazioni.

Nel corso degli ultimi quattro mesi era stato mostrato probabile --sulla base del lavoro svolto da Joliot in Francia, e da Fermi e Szilard in America—che si potesse generare in una grande massa di uranio una reazione nucleare a catena, attraverso la quale raggiungere una potenza altissima e produrre elementi del tipo del radio. Ora sembra praticamente certo che ciò possa essere conseguito nell'immediato futuro.

Questo nuovo fenomeno condurrebbe anche alla costruzione di bombe, e si può pensare – benchè ciò sia molto meno sicuro—che vengano così costruite bombe di nuovo tipo estremamente potenti. Una singola bomba di questo tipo, trasportata da una nave e fatta esplodere in un porto, potrebbe facilmente distruggere l'intero porto e parte della regione adiacente. Però una tale bomba potrebbe risultare troppo pesante per essere trasportata da un aereo.

Gli Stati Uniti posseggono solo quantità molto modeste, e di basso tenore, di minerali di uranio. C'è del buon minerale in Canada e nella ex-Cecoslovacchia; mentre la maggiore fonte di uranio è il Congo Belga.

A cagione di questa situazione, Lei potrebbe considerare l'opportunità di mantenere un contatto permanente tra il Suo Esecutivo e il gruppo di fisici che lavorano sulle reazioni a catena in America. Una possibile via per ottenere questo potrebbe essere per Lei quella di affidare detto incarico a persona di Sua fiducia e che magari agisca in maniera confidenziale. Il suo compito dovrebbe includere:

a) prendere contatti coi Ministeri del Governo, tenerli informati dei prossimi sviluppi e formulare raccomandazioni per l'intervento del Governo, prestando particolare attenzione al problema di assicurare un rifornimento di minerale di uranio agli Stati Uniti;

b) accelerare il lavoro sperimentale, che attualmente viene portato avanti entro i limiti dei fondi dei laboratori universitari, fornendo fondi, se tali fondi fossero necessari, attraverso i suoi contatti con privati disposti a contribuire alla causa, e magari ottenendo la cooperazione di laboratori industriali dotati dell'attrezzatura necessaria.

Mi risulta che la Germania ha interrotto la vendita di uranio che le viene dalle miniere della Cecoslovacchia che lei ha occupato. Che la Germania abbia adottato questo tempestivo provvedimento potrebbe forse essere compreso sulla base del fatto che il figlio del suo Sotto-Segretario di Stato, von Weizsaecker, lavora presso l'Istituto Kaiser-Wilhelm di Berlino, ove parte del lavoro eseguito in America sull'uranio è in corso di ripetizione.

<div style="text-align: center;">
Coi migliori saluti,
Albert Einstein
</div>

*************************************************************

# Breve Inserto Storico
# sul periodo 1933-1945

La Germania era uscita sconfitta e impoverita dalla prima guerra mondiale. Rivolte dei militari e di intere città avevano obbligato l'imperatore Guglielmo II ad abdicare. Presto era nata la Repubblica di Weimer, socialdemocratica, che aveva spento le insurrezioni di tipo comunista. Ma il trattato di pace del 1919, punitivo per la Germania, sancisce perdite territoriali e delle colonie a favore degli Stati confinanti; impone limitazioni alle forze armate e riparazioni per i danni di guerra; alimenta sentimenti nazionalistici di rivalsa e aggrava la crisi economica e sociale tedesca, aprendo la strada ad anni di gravissima inflazione e miseria. La Germania si apre alla Russia, finora isolata, e nasce una collaborazione proficua per entrambe le nazioni; la Germania ottiene la cancellazione del suo debito di guerra con la Russia; la situazione del popolo comincia a migliorare. Ma il "Dicktat" a cui la Germania è soggetta sembra fatto apposta per impedirne lo sviluppo. La Germania non riesce a pagare i debiti a Francia e Belgio, e questi occupano (1923) il bacino della Ruhr, che produceva i 4/5 del carbone tedesco. I minatori fanno resistenza passiva, ma ciò porta la crisi socio-economica della Germania al culmine. Un intervento degli Stati Uniti migliora la situazione, ma la crisi del 1929 annulla presto gli effetti positivi dei piani americani Dawes e Young per le riparazioni tedesche dei debiti di guerra, e si riaccende la tensione sociale in Germania, ove i disoccupati salgono a 6 milioni. Ciò facilita ovviamente la scalata del potere dei Nazionalsocialisti hitleriani, che iniziano ad avere successi elettorali nel 1930 e 1932.

  Il 30 gennaio 1933 il presidente della repubblica di Weimer, von Hindenburg, nomina cancelliere (cioè Primo Ministro) Hitler, il quale ottiene i pieni poteri il 23 marzo dello stesso anno. Finisce la Repubblica di Weimer e inizia il III Reich. Comincia la fuga dalla Germania degli anti-nazisti, tra i quali Albert Einstein. Alla morte del presidente von Hindenburk, Hitler assume anche la carica di capo dello stato. Con le leggi
del 1935 si intensifica la persecuzione antisemita. Nel 1936, dopo la Guerra di Spagna, viene stipulato l'Asse Roma-Berlino, mentre la Germania firma il patto Anti-Kominter col Giappone. Nel 1938 Hitler annette l'Austria ("Anschluss"); mentre Gran Bretagna, Francia e Italia danno via libera alla Germania per l'occupazione del territorio dei Sudeti. Nel 1939, Hitler occupa la Cecoslovacchia, stringe un "patto d'acciaio" con l'Italia, e prende ulteriori accordi col Giappone (interessato a divenire il Paese asiatico egemone, cosicché l'Asse Roma-Berlino diventerà triangolare, includendo Tokio, agli inizi del 1941). Infine, subito dopo, firma un patto di non aggressione con la Russia (con l'accordo segreto di spartirsi la Polonia). La guerra era oramai pronta, e il 2 agosto del 1939 Einstein firma la sua nota lettera a Roosevelt.

La richiesta da parte di Hitler di un corridoio attraverso la Polonia verso il porto di Danzica (che sarebbe stato l'unico porto marittimo della Germania), richiesta alla quale si oppongono Inghilterra e Francia, fa scatenare la II Guerra Mondiale. Il 1^o settembre del 1939 Hitler ordina una guerra-lampo contro la Polonia (che in effetti, come tutti si aspettavano, non viene difesa da nessuno). Seguono, ad intervalli, altre guerre-lampo, come quella contro la Francia, attraversando il neutrale Belgio così da superare banalmente la Linea Maginot. Ma dopo il lungo periodo di veloci vittorie, l'Inghilterra comincia a combattere; e simultaneamente Russia e Stati Uniti fortificano l'industria bellica (il che, tra parentesi, risolve i gravi problemi economici ancora presenti negli USA).

Quando nel 1941 la Germania invade la Russia, il suo destino è segnato, e la sconfitta viene accelerata dalla dichiarazione di guerra, da parte
del Giappone (07.12.41), e successivamente di Germania e Italia (11.12.41), agli Stati Uniti. L'Italia firma un armistizio l'8 settembre 1943. Il 7 maggio 1945 la Germania è costretta a una resa senza condizioni, e Hitler si suicida. Continua la guerra solo contro il Giappone.

A questo punto, il 6 e 9 agosto del 1945 gli Stati Uniti sganciano due bombe atomiche (sperimentando sia quella ad Uranio arricchito, sia quella a Plutonio) rispettivamente su Hiroshima e Nagasaki; così che anche il Giappone firma una resa incondizionata il 2 settembre 1945.

# *La responsabilità degli scienziati, E.Amaldi, L.Sciascia, e E.Majorana*

Al Progetto Manhattam per la costruzione, a Los Alamos (New Mexico, USA) della bomba A partecipò, tra gli altri, Emilio Segré, uno dei "ragazzi di via Panisperna", ovvero uno dei collaboratore di Fermi a Roma (che otterrà il premio Nobel negli USA per la scoperta dell'antiprotone).

Segré, parlando con **Leonardo Sciascia**, si vanterà di avere partecipato alla costruzione della Bomba: e Sciascia, indignato, dedicherà un suo libro (*La scomparsa di Majorana*) alla questione della responsabilità della scienza: mitizzando la figura di Ettore Majorana (altro "ragazzo di via Panisperna", e fisico teorico insuperabile) quale simbolo dello scienziato che fugge di fronte al rischio che i politici sfruttino le loro scoperte a fini militari.

*Sciascia ha poi dato un'importanza via via crescente a questa sua preoccupazione per le applicazioni negative che possono ricevere le scoperte scientifiche, e quindi al suo relativo saggio sulla scomparsa di Ettore Majorana. Perché tanto interesse, e duraturo nel tempo, da parte di Leonardo Sciascia per questo personaggio e per le vicende della sua scomparsa avvenuta nel lontano 1938? L'abbiamo già intuito, ma vogliamo aggiungere qualche particolare.*

In un altro libro di Sciascia, *Fatti Diversi di Storia Letteraria e Civile* (Sellerio, Palermo), possiamo leggere quanto da lui scritto a proposito del proprio racconto su Majorana:

*«L'avevo scritto, questo racconto, nella memoria che avevo della scomparsa e su documenti che, per tramite del professor Recami, ero riuscito ad avere, dopo aver casualmente sentito un fisico* [Emilio Segré] *parlare con soddisfazione, ed entusiasmo persino, delle bombe che avevano distrutto*

*Hiroshima e Nagasaki. Per indignazione, dunque; e tra documenti e immaginazione, i documenti aiutando a rendere probante l'immaginazione, avevo fatto di Majorana il simbolo dell'uomo di scienza che rifiuta di immettersi in quella prospettiva di morte cui altri, con disinvoltura a dir poco, si erano avviati».*

                                 Albert Einstein
                                 Old Grove Rd.
                                 Nassau Point
                                 Peconic, Long Island

                                 August 2nd 1939

F.D. Roosevelt
President of the United States
White House
Washington, D.C.

Sir:

    Some recent work by E.Fermi and L. Szilard, which has been communicated to me in manuscript, leads me to expect that the element uranium may be turned into a new and important source of energy in the immediate future. Certain aspects of the situation which has arisen seem to call for watchfulness and, if necessary, quick action on the part of the Administration. I believe therefore that it is my duty to bring to your attention the following facts and recommendations:

    In the course of the last four months it has been made probable - through the work of Joliot in France as well as Fermi and Szilard in America - that it may become possible to set up a nuclear chain reaction in a large mass of uranium, by which vast amounts of power and large quantities of new radium-like elements would be generated. Now it appears almost certain that this could be achieved in the immediate future.

    This new phenomenon would also lead to the construction of bombs, and it is conceivable - though much less certain - that extremely powerful bombs of a new type may thus be constructed. A single bomb of this type, carried by boat and exploded in a port, might very well destroy the whole port together with some of the surrounding territory. However, such bombs might very well prove to be too heavy for transportation by air.

                        -2-

> The United States has only very poor ores of uranium in moderate quantities. There is some good ore in Canada and the former Czechoslovak while the most important source of uranium is Belgian Congo.
>
> In view of the situation you may think it desirable to have more permanent contact maintained between the Administration and the group of physicists working on chain reactions in America. One possible way of achieving this might be for you to entrust with this task a person who has your confidence and who could perhaps serve in an unofficial capacity. His task might comprise the following:
>
> a) to approach Government Departments, keep them informed of the further development, and put forward recommendations for Government act giving particular attention to the problem of securing a supply of uranium ore for the United States;
>
> b) to speed up the experimental work, which is at present being carried on within the limits of the budgets of University laboratories, by providing funds, if such funds be required, through his contacts with y private persons who are willing to make contributions for this cause, and perhaps also by obtaining the co-operation of industrial laboratories which have the necessary equipment.
>
> I understand that Germany has actually stopped the sale of uranium from the Czechoslovakian mines which she has taken over. That she shoul have taken such early action might perhaps be understood on the ground that the son of the German Under-Secretary of State, von Weizsäcker, is attached to the Kaiser-Wilhelm-Institut in Berlin where some of the American work on uranium is now being repeated.
>
> Yours very truly,
> 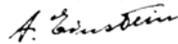
> (Albert Einstein)

*Fig.7*

Da questo brano si deduce la vera ragione dell'interesse costante di Sciascia per l'argomento ivi toccato: ovvero, per la *vexata questio* della responsabilità della scienza e degli scienziati. Premettiamo alcune parole di cronaca.

Anzitutto, l'incontro di Sciascia con Segrè avvenne a pranzo, in Svizzera, presente Moravia, il quale non si trattenne dal dare qualche gomitata sotto il tavolo a Leonardo quando Segrè cominciò a vantare il suo ruolo nella costruzione della bomba A (e, come vedremo, a Segré non mancava una sua piccola dose di ragioni).

L'agrigentino Sciascia scelse, in contrapposizione a Segré, il grande conterraneo (catanese) Ettore Majorana –paragonato da Enrico Fermi a geni come Galilei e Newton-- quale esempio, si diceva, dello scienziato che, di fronte al pericolo che le proprie scoperte possano essere usate a fin di male dal potere economico e politico, rinuncia a renderle note, e si ritira nell'ombra. Forse Leonardo esagerò in questa simbolica contrapposizione; d'altra parte, come lui stesso scrisse, e come abbiamo già menzionato, il suo racconto è «un misto di storia e di invenzione»... [Aggiungiamo, tra parentesi, che nel gruppo di Fermi ci fu davvero chi, sapendo di Los Alamos e della costruzione della bomba, abbandonò la fisica: il grande sperimentale Franco Rasetti. Abbandonata la fisica, divenne un paleontologo di rinomanza internazionale; e, già avanti negli anni, passò poi alla botanica, divenendo uno dei maggiori esperti mondiali di orchidacee].

Esaminando il saggio di Sciascia, si può verificare la profondità dell'intuito psicologico di Leonardo, e della sua alta capacità di *intus legere*. Leggendo tra le righe, appunto, e in mezzo alle carte, Leonardo ha saputo intuire che il grande fisico teorico Werner Heisenberg e gli altri scienziati tedeschi *non vollero* accingersi alla costruzione di una bomba atomica: commentando che gli schiavi (di Hitler) si comportarono da liberi, mentre i liberi (gli Americani) si comportarono da schiavi... Anche a questa tesi, che raccoglieva ben pochi sostenitori, Sciascia ci teneva; e la sua conferma è arrivata, eclatante, nel 1993, dopo la morte di Sciascia, quando il *British Intelligence Service* ha tolto il segreto ai "Farm Hall Transcripts". Spieghiamoci. Tra il giugno e il dicembre del 1945 (un periodo che comprese il bombardamento di Hiroshima del 6 agosto), per 24 settimane, dieci tra i più importanti scienziati tedeschi [tra i quali Heisenberg, von Weizsaecker, Otto Hahn, Walther Gerlach, Max von Laue,...] furono tenuti prigionieri nella Farm Hall, presso Cambridge, in Inghilterra, e le loro conversazioni furono di nascosto registrate dal servizio segreto britannico. La traduzione inglese di tali conversazioni (in particolare delle reazioni dei reclusi quando giunse la notizia di Hiroshima e Nagasaki) è apparsa in stampa nel **1993** nel volume *Operation Epsilon: The Farm Hall Transcripts* (**I.O.P. Pub.; Bristol, UK**). Dalle suddette trascrizioni risulta evidente che su dieci, solo uno scienziato tedesco (non Heisenberg!) avrebbe voluto, potendo, contribuire alla costruzione della bomba A tedesca.

Sciascia ebbe poi l'impressione di un latente antagonismo tra Ettore Majorana ed Enrico Fermi, un antagonismo negato da tutti i colleghi e amici di Ettore, ma che, col senso di poi (Ettore abbandonò non solo la famiglia, ma anche il gruppo romano guidato da Fermi) potrebbe contenere un qualche briciolo di verità. Tale presa di posizione di Sciascia generò una vivace polemica**(\*)** tra Leonardo e i fisici, in particolare Edoardo Amaldi. Questo ci interessa perché presto la polemica divenne aspra, tanto che Sciascia arrivò a scrivere (su "La Stampa" della vigilia di Natale del 1975) che "si vive come cani per colpa della scienza": in ciò associandosi a una tradizione di pensiero tipicamente italiana e non molto nobile.

Cosa voleva dire Sciascia? Crediamo che lui sapesse che non esistono la scienza o la poesia, ma solo scienziati e poeti; e che le colpe ricadrebbero semmai su (alcuni) scienziati. Crediamo inoltre che Sciascia sarebbe stato d'accordo nell'affermare che la colpa dell'esistenza della sedia elettrica non è di Alessandro Volta; così come la colpa di una rapina a mano armata non è dell'inventore del coltello.

Comunque Sciascia ha voluto rinverdire un problema vecchio come il mondo. È nato con Prometeo, quando l'uomo ha incominciato a controllare il fuoco. È un problema che ha sentito Alfred Nobel quando, avendo costruito la dinamite (che allevia la fatica delle braccia dell'uomo, ma può divenire arma bellica), creò il Premio Nobel, quasi come atto di espiazione.

Ma rileggiamo prima una interessante affermazione di Sciascia, da lui scrittaci in una sua lettera del 27 gennaio '76: «*Voglio fare presente che per me l'espressione "rifiuto della scienza" vale "rifiuto della scienza a un certo punto di fronte a certe ricerche, a certe scoperte". E cioè rifiuto da parte degli scienziati stessi*».

Abbandoniamo però la cronaca e torniamo al tema principale che ha ispirato Sciascia: il problema della responsabilità degli uomini di scienza.

Il problema delle scoperte e invenzioni umane (fosse anche solo quella del martello), le quali ammettono applicazioni positive e negative è, dicevamo, un problema antico; che ha sentito ad esempio anche Pierre Curie, il quale, nel ricevere il premio Nobel per la mitica scoperta del *radium,* ebbe a dichiarare: « *Si può pensare che in mani criminali il* radium *possa divenire molto pericoloso, e ci si può chiedere se l'umanità tragga profitto dalla conoscenza dei segreti della natura. L'esempio della scoperta di Nobel [anche Curie lo cita] è caratteristico: Gli esplosivi permettono all'uomo di compiere opere mirabili. Essi sono però anche un mezzo di distruzione in mano ai grandi criminali che spingono i popoli alla guerra. Ma io --conclude Curie-- sono tra quelli che credono che l'umanità trarrà più bene che male dalle nuove scoperte* ».

Si può proporre un'altra considerazione. La costruzione di strumenti è caratteristica *ineliminabile* dell'uomo. Mentre molti animali nell'evoluzione biologica, avendo bisogno per esempio di mascelle più robuste, sviluppano i muscoli della mandibola, l'uomo non fa altrettanto, ma costruisce a partire da una pietra un coltello di selce. E, se ha bisogno di un braccio più robusto, si limita ad usare un randello: fabbrica, in altre parole, prolungamenti artificiali dei propri organi. E' *inevitabile* che l'uomo costruisca randelli, e martelli, anche se questi possono essere usati contro i propri simili.

E' forse un problema solo degli scienziati quello del controllo, e dell'uso a fin di bene, delle scoperte e delle invenzioni umane?

Precisiamo alcuni termini della questione. Lo scienziato vero è quello che fa ricerca solo per amore della conoscenza: per scoprire qualcosa degli eleganti segreti della natura che ci circonda. E' il tecnologo che si occupa invece delle eventuali applicazioni dei risultati della ricerca scientifica (anche se lo stesso individuo, in quanto uomo, a un certo punto può smettere i panni dello scienziato per cambiare mestiere, e assumere quelli del tecnologo). Eventuali "colpe" dovrebbero essere attribuite, semmai, ai tecnologi. Ma il tecnologo stesso può giungere alla costruzione, al massimo, di *un unico* prototipo: *una* primitiva automobile a vapore, ad esempio. E' poi l'intervento del potere economico e politico a determinare la produzione, o meno, di innumerevoli copie del prototipo. I poteri da controllare, pertanto, sono quello economico e politico, che purtroppo si ispirano quasi esclusivamente al tornaconto, per conseguire il quale scatenano guerre economiche e guerre vere. E' ovvio pertanto che questo controllo non può essere demandato alle povere forze degli scienziati, e neppure a quelle dei soli tecnologi: ma esso è compito e dovere di *tutti* i cittadini.

Possiamo rispondere agli stimoli di Sciascia riconoscendo che *anche* gli scienziati devono porsi i problemi che *tutti noi* dobbiamo porci; e ricordandoci della responsabilità che noi tutti abbiamo di fronte all'uso che si fa, nel bene e nel male, delle conquiste del "progresso".

Abbiamo visto come in realtà non sia soltanto lo scienziato, o non sia soprattutto lo scienziato, ad avere le responsabilità di cui parliamo. Poniamoci ora una domanda, questa volta di tipo scientifico-biologico: come mai l'uomo, fra tutti gli animali, è quello che apparentemente è il più feroce coi propri simili? Perché li attacca e tortura, mentre la maggior parte degli animali non si comporta in tal modo? Perchè è essenziale, quindi, *controllare* il potere? Una ragione biologica c'è; ed è la seguente.
Gli animali che nascono con mezzi di offesa scadenti e deboli (come gli uomini, con i loro poveri denti e unghie) non ricevono in dono dalla natura l'istinto del "cavalierismo" verso i propri simili; mentre gli animali dotati di mezzi di offesa potenti –come lupi o tigri-- posseggono di necessità l'istinto del rispetto intraspecifico: altrimenti la loro specie si sarebbe già estinta! Gli agnelli, per esempio, non hanno denti poderosi, non hanno artigli, e quindi la natura non li ha istintivamente dotati di rispetto per i propri simili; tanto che, trovandosi due agnelli sul ciglio di un burrone, può ben avvenire che uno spinga l'altro giù dal dirupo. Analogamente per due piccioni: essi pure posseggono deboli mezzi di offesa; quando eseguono le loro battaglie mimiche per conquistare il predominio su un territorio, ad un certo punto uno dei due si riconosce perdente, e se ne vola via: e basta. Ma se si prendono due piccioni maschi e li si mettono in una unica gabbia, il vincitore torturerà a morte il perdente... Quando invece sono due lupi a
recitare la mimica della loro battaglia (una mimica dalla quale è forse nata la nostra danza) per la conquista del predominio sul branco, a un certo punto uno dei due lupi riconosce la propria inferiorità: questi allora si arrende, e offre il collo, esponendo la giugolare, al vincitore. Il vincitore, nonostante possa dimostrare una gran voglia di azzannare il soccombente, in realtà è costretto dall'istinto a comportarsi da cavaliere: il primo si arrende e il secondo invariabilmente accetta la sua resa e gli concede la vita.

Noi uomini non abbiamo ricevuto *in dono* dalla natura l'istinto del rispetto del prossimo. Però abbiamo poi costruito coltelli, fucili, le bombe atomiche... Che cosa occorre allora? Che il rispetto dei nostri simili ce lo dobbiamo guadagnare con ogni sforzo verso la maturazione morale, la quale *deve* crescere man mano che le nostre capacità di offesa artificiali aumentano. Si potrebbe dire che Dio non ci ha voluto fare un tale tipo di regalo, affinché ce lo si dovesse guadagnare, attraverso uno sforzo libero per lo sviluppo della nostra coscienza morale. Noi ora abbiamo in mano mezzi potentissimi, come aerei da guerra, bombe micidiali, armi chimiche e batteriologiche; e bombe all'Idrogeno: oggi basta schiacciare un bottone per uccidere un milione di uomini. *Quindi* abbiamo l'obbligo di sforzarci –per la sopravvivenza della nostra specie— al fine di conseguire una grande maturazione morale. Maturazione che deve certamente aver luogo negli scienziati, nei fisici, ora diremmo nei biologi; ma che dobbiamo avere tutti, perché è un compito che l'intera l'umanità deve affrontare: solo l'unione di intenti di tutti i cittadini del mondo può imporre ai veri potenti di perseguire fini di pace.

Prima di concludere, approfittiamo della nostra citazione di Sciascia, che ha cominciato il suo impegno civile quale educatore, riaffermando che la preparazione più importante che noi giovani dobbiamo chiedere alla scuola per la vita è l'attuazione delle nostre potenzialità ereditarie, la maturazione delle nostre facoltà morali, intellettuali e cognitive: e non tanto la preparazione a svolgere un mestiere. Probabilmente la cultura di base che noi abbiamo (o avevamo?) in Italia, e più in generale in Europa, è insostituibile. La mancanza di una forte cultura di base è deleteria; è importante «perdere tempo» nello studio: perché sono proprio le cose che apparentemente non servono a niente, come la letteratura, la poesia, la filosofia, la scienza intesa come conoscenza del mondo, che formano la mente.

*Il rapporto tra intellettuali e potere*
Abbiamo visto come il dovere di controllare il Potere sia di tutti i cittadini. In particolare, aggiungiamo, di tutti gli Intellettuali. Siamo così arrivati alla importante questione del rapporto Intellettuali e Potere.

Spesso gli intellettuali, per ricevere benefici dal Potere, lo sostengono anche quando esso non opera per il bene della società.

A questo punto diviene spontaneo ricordare, ancora una volta, coloro che subirono la repressione del Potere per essersi opposti ad attitudini vessatorie o liberticide o troppo conservatrici. Abbiamo già considerato il caso di Seneca, e di Einstein, che dovettero vivere a lungo in esilio. Limitandoci alla letteratura inglese, autori come Oscar Wilde o Ezra Pound, ad esempio, sono stati anche in prigione (o in manicomio) per avuto il coraggio di affermare le proprie tendenze omosessuali o politiche. E Virginia Woolf soffrì tutta la vita per la emarginazione delle donne nella sua società, cosa che contribuì al suo suicidio finale.

*(Erasmo Umberto M. Recami,*
*Umberto Victor G. Recami,*
*& Erasmo Recami)*

**SEGUE UNA *APPENDICE* "SELF-CONTAINED"**
RIGUARDANTE IL SOLO TEMA
" MAJORANA, SCIASCIA, AMALDI, E LA RESPONSABILITA'
DEGLI SCIENZIATI " **:**
==================================================

\* \* \* \* \* \* \* \*

# APPENDICE / A SELF-CONTAINED APPENDIX

Articolo separato (AUTOSUFFICIENTE) sulla **sola** questione:

## *Majorana, Sciascia, Amaldi e la responsabilità degli intellettuali*

==================================================

[January 2006)]

Lo scorso anno, 2005, ha visto il trentennale della pubblicazione del saggio di Leonardo Sciascia su "La scomparsa di Ettore Majorana" (tanto che la Fondazione Sciascia vi ha dedicato in novembre un convegno a Racalmuto, paese natale di Sciascia), mentre il corrente anno 2006 è il centenario della nascita di Majorana: ovvero, con ogni probabilità, della nascita del maggior fisico teorico italiano del secolo trascorso. Vale la pena tornare a valutare criticamente i temi sostenuti o toccati da Sciascia nel suo libro e negli scritti giornalistici successivi; temi che riguardano da vicino la questione della responsabilità dell'intellettuale nei confronti dell'uso che può essere fatto delle conquiste della tecnologia.

Invero, Leonardo Sciascia ha dato un'importanza via via crescente al suo detto saggio sulla scomparsa di Majorana. In un'intervista, segnalatami a suo tempo dall'amico Enzo Vitale e in cui a Leonardo Sciascia veniva richiesto di dire quale fosse tra i suoi libri quello che a lui più piacesse, Leonardo rispose: «Fino a qualche anno fa, avrei detto *Morte di un Inquisitore,* ora invece rispondo *La Scomparsa di Ettore Majorana.*». Sciascia-detective non poteva non essere affascinato da quel giallo di alto livello culturale, quale è la vicenda relativa alla scomparsa dell'eccelso fisico teorico. Ma perché tanto interesse, e duraturo nel tempo, da parte di Sciascia per questo personaggio e per le vicende della sua scomparsa avvenuta nel lontano 1938 ?

In un libro di Sciascia, *Fatti Diversi di Storia Letteraria e Civile* (Sellerio, Palermo), possiamo rileggere quanto da lui scritto in origine su "La Stampa" di Torino per commentare la tarda pubblicazione, da parte di Emilio Segrè, di una discussa lettera indirizzata a quest'ultimo dal Majorana nel 1933. A proposito del proprio racconto "misto di storia e d'invenzione", Sciascia aveva dichiarato:

*«L'avevo scritto, questo racconto, nella memoria che avevo della scomparsa e su documenti che, per tramite del professor Recami, ero riuscito ad avere, dopo aver casualmente sentito un fisico parlare con soddisfazione [il titolo del pezzo giornalistico, qui, è "Majorana e Segré"], ed entusiasmo persino, delle bombe che avevano distrutto Hiroshima e Nagasaki. Per indignazione, dunque; e tra documenti e immaginazione, i documenti aiutando a rendere probante l'immaginazione, avevo fatto di Majorana il simbolo dell'uomo di scienza che rifiuta di immettersi in quella prospettiva di morte cui altri, con disinvoltura a dir poco, si erano avviati ».*

Questo brano già ci rivela la vera ragione dell'interesse costante di Sciascia per l'argomento ivi toccato: ovvero, per la vexata questio della responsabilità della scienza e degli scienziati. Concediamoci due parole di cronaca. Anzitutto, l'incontro di Sciascia con Emilio Segrè (già membro del famoso gruppo romano dei "ragazzi di via Panisperna, guidato da Enrico fermi) avvenne in Svizzera, presente Moravia, il quale non si peritò di dare qualche gomitata sotto il tavolo a Leonardo quando Segrè cominciò a vantare il suo ruolo nella costruzione della bomba A (e, come vedremo, a Segré non mancava una sua piccola dose di ragioni). Successivamente, nel 1972 un amico comune, il professor Carapezza di Palermo, telefonò al sottoscritto informandolo che Leonardo Sciascia si era deciso a scrivere un'opera su Ettore Majorana e mandava a chiedere se pure noi fossimo interessati a scrivere un libro sull'argomento; altrimenti avrebbe avuto il piacere di conoscere i documenti in mio possesso. Chi scrive aveva, sì, l'intenzione di condurre in porto, già allora, il proprio volume su Il Caso Majorana: ma fu ben felice di cedere il passo, e di rimandare il proprio libro di una decina d'anni: per la valorizzazione che la penna del famoso scrittore avrebbe certamente apportato alla figura di Ettore Majorana, allora nota fuori d'Italia quasi soltanto tra i fisici. E, accompagnando Leonardo a Roma, convincemmo nel 1972 la sorella di Ettore, Maria, inizialmente un poco restia, a concedergli copia di parte dei documenti esistenti (documenti, tra l'altro, tutti rinvenuti o raccolti da chi scrive, sempre in pieno accordo con l'indimenticabile Maria Majorana).

L'agrigentino Sciascia scelse, come contraltare di Segré, il grande conterraneo (catanese) Ettore Majorana –-paragonato da Enrico Fermi a geni come Galilei e Newton-- quale esempio dello scienziato che, di fronte al pericolo che le proprie scoperte possano venire usate a fin di male dal Potere , rinuncia a renderle note, e si ritira nell'ombra. Tale simbolica contrapposizione fu essenzialmente una finzione letteraria; d'altra parte, come Sciascia medesimo scrisse, e come abbiamo già menzionato, il suo racconto è «un misto di storia e di invenzione»: così che, confondendo volontariamente l'essere col dover essere, Sciascia arrivò ad attribuire ad Ettore anche qualità, interessi e decisioni funzionali alla trasformazione della vicenda di Majorana in emblema del comportamento dello scienziato "buono" di fronte ai problemi posti dal progresso scientifico. Aggiungiamo, tra parentesi, che nel gruppo di Fermi ci fu davvero chi, sapendo di Los Alamos e della costruzione della bomba, abbandonò la fisica: il grande sperimentale Franco Rasetti. Tralasciata la fisica, divenne un paleontologo di rinomanza internazionale; e, già avanti negli anni, passò poi alla botanica, divenendo alfine uno dei maggiori esperti mondiali di orchidacee.

Sciascia, comunque, indugiò per qualche anno; finchè, accogliendo un invito scritto, nella primavera del 1975 noi lo si raggiunse nella sua casa di Racalmuto, in contrada Noce, e si contribuì a convincerlo a comporre quell'estate, finalmente, il suo libretto sulla scomparsa di Majorana.

Esaminando il saggio di Sciascia, si può verificare più di una volta la capacità di *intus legere*, che accompagna l'arte meditata della parola di cui Leonardo Sciascia imbeve i suoi racconti. Leggendo tra le righe, appunto, e in mezzo alle carte, Leonardo seppe intuire alcuni aspetti che sembrano rispondere a verità, come la scoperta di ulteriori documenti negli anni successivi ha parzialmente confermato. Significativa, ad esempio, è la circostanza che Leonardo sostenne che Werner Heisenberg e gli altri scienziati tedeschi *non vollero* accingersi alla costruzione di una bomba atomica: commentando –come noto-- che gli schiavi (di Hitler) si comportarono da liberi... A questa tesi, che raccoglieva ben pochi sostenitori, Sciascia ci teneva; e la conferma di essa è arrivata, eclatante, agli inizi degli anni '90, dopo la dipartita di Sciascia, quando il British Intelligence Service ha tolto il segreto ai "Farm Hall Transcripts". Spieghiamoci. Tra il giugno e il dicembre del 1945 (un periodo che comprese il bombardamento di Hiroshima del 6 agosto), per 24 settimane, dieci tra i più importanti scienziati tedeschi [tra cui Heisenberg, von Weizsaecker, Otto Hahn, Walther Gerlach, Max von Laue] furono tenuti prigionieri nella Farm Hall, presso Cambridge, UK, e le loro conversazioni furono registrate dal servizio segreto britannico a loro insaputa. La traduzione inglese di tali conversazioni (in particolare delle reazioni dei reclusi quando giunse la notizia di Hiroshima e Nagasaki) è apparsa in istampa nel **1993** nel volume *Operation Epsilon: The Farm Hall Transcripts* (**I.O.P. Pub.; Bristol, UK**). Dalle suddette trascrizioni risulta evidente che, su dieci, solo uno scienziato tedesco (non Heisenberg!) avrebbe voluto, potendo, contribuire alla costruzione della bomba A tedesca.

Significativo è pure il fatto che Sciascia si convinse presto che la scomparsa di Ettore si riferiva a una fuga e non ad un suicidio: ipotesi che sembra la più probabile alla luce dei documenti, pur non decisivi, da noi successivamente rintracciati.

Ebbe poi l'impressione di un latente antagonismo tra Ettore Majorana ed Enrico Fermi, un antagonismo negato da tutti i colleghi e amici di Ettore, ma che, col senno di poi (Ettore abbandonò non solo la famiglia, ma anche il gruppo guidato da Fermi) potrebbe contenere un qualche briciolo di verità. Tale presa di posizione di Sciascia generò, come molti ricorderanno, una vivace polemica tra Leonardo e i fisici, in particolare Edoardo Amaldi; polemica nella quale il sottoscritto prese le parti più di Amaldi che di Sciascia. La polemica riguardò inizialmente quasi solo la questione della partecipazione del Majorana al famoso concorso universitario per la fisica teorica nel 1937 (partecipazione voluta dal gruppo di Fermi –come anche a noi parrebbe-- o decisa da Ettore in contrasto coi colleghi?): ed essa ci vide nella singolare situazione di amici di entrambi i maggiori contendenti, i quali entrambi si sfogavano anche con l'invio di acuminate epistole –l'un contro l'altro armati— al sottoscritto.**(\*)** La polemica presto divenne aspra, tanto che Sciascia arrivò a scrivere (su "La Stampa" della vigilia di Natale del 1975) che "si vive come cani per colpa della scienza": in ciò associandosi un po' pedissequamente a una tradizione di pensiero tipicamente italiana e non molto nobile, che annovera comunque nomi quali il Vico e Benedetto Croce.

Cosa voleva dire Sciascia? Crediamo che lui sapesse che non esistono la scienza o la poesia, ma solo scienziati e poeti; e che le colpe ricadrebbero semmai su (alcuni) scienziati. Crediamo che sapesse, per di più, che, se un poeta o un filosofo pessimisti offrono a un infelice la goccia che lo decide a commettere suicidio, vere colpe non si possano attribuire a quel filosofo o poeta…

Parlando con Sciascia, si era d'accordo nel dire che la colpa dell'esistenza della sedia elettrica non è affibbiabile ad Alessandro Volta; così come la colpa di una rapina a mano armata non è dell'inventore del coltello. Comunque Sciascia ha voluto rinverdire un ricorrente problema, già riproposto in anni non lontani, e in maniera più *soft,* per esempio da Duerrematt, e a proposito del quale proporremo alcune considerazioni: basate in parte sulla constatazione che il problema della potenzialità distruttiva degli strumenti che l'uomo costruisce è vecchio come il mondo. È nato con Prometeo, quando l'uomo ha incominciato a controllare il fuoco. È un problema che ha sentito Alfred Nobel quando, avendo costruito la dinamite (che allevia la fatica delle braccia dell'uomo, ma può divenire arma bellica), creò il premio Nobel, quasi come atto di espiazione.

Ma rileggiamo prima alcune affermazioni di Sciascia, e di Amaldi. In una lettera del 3 dicembre del '75 Amaldi ci dice: «Credo che avrà anche visto l'intervista di Sciascia sul Giornale di Sicilia del 9 novembre del '75. Quì finalmente viene fuori la vera posizione di Sciascia, ossia quella classica in Italia di Croce, di Gentile: la scienza non fa parte della cultura». E Leonardo pochi giorni dopo, il 9 dicembre '75, ci scrive, riferendosi al suo lungo articolo inviato a La Stampa e apparso, come detto, il 24 dicembre del 1975, commentando: «Naturalmente questa è l'ultima volta che scendo in polemica, ed è il caso di dire scendo perché la polemica di Amaldi è piuttosto bassa». In un'altra lettera, il 27 gennaio '76, Leonardo aggiunge infine una affermazione interessante: «Voglio soltanto fare presente che per me l'espressione "rifiuto della scienza" vale "rifiuto della scienza a un certo punto di fronte a certe ricerche, a certe scoperte". E cioè rifiuto da parte degli scienziati stessi».

Abbandoniamo la cronaca, e torniamo al nostro tema principale: il problema della responsabilità degli uomini di scienza.

Premettiamo che Sciascia è uno dei pochissimi scrittori che abbiano parlato di uno scienziato attribuendogli ricchezza e spessore umani e non povertà spirituale; e questo lascia ben sperare per la soluzione del problema annoso delle due culture.

Il problema delle scoperte e invenzioni umane (fosse solo quella del martello) che ammettono applicazioni positive e negative è, dicevamo, un dilemma antico; che ha sentito anche Pierre Curie (il fisico, consorte di Madame Curie), il quale, nel ricevere il premio Nobel per la mitica scoperta del *radium,* ebbe a dichiarare: «Si può pensare che in mani criminali il *radium* possa divenire molto pericoloso, e ci si può chiedere se l'umanità tragga profitto dalla conoscenza dei segreti della natura. L'esempio della scoperta di Nobel [anche Curie lo cita] è caratteristico: Gli esplosivi permettono all'uomo di compiere opere mirabili. Essi sono però anche un mezzo di distruzione in mano ai grandi criminali che spingono i popoli alla guerra. Ma io --conclude Curie-- sono tra quelli che credono che l'umanità trarrà più bene che male dalle nuove scoperte».

Suggeriamo un'altra considerazione. La costruzione di strumenti è caratteristica *ineliminabile* dell'uomo. Mentre molti animali nell' evoluzione biologica, avendo bisogno per esempio di mascelle più robuste, sviluppano i muscoli della mandibola, l'uomo non fa altrettanto: ma costruisce a partire dalla pietra un coltello di selce. E, se ha bisogno di un braccio più robusto, si limita ad usare un randello; fabbrica, in altre parole, prolungamenti artificiali dei propri organi. E' inevitabile che l'uomo costruisca randelli, e martelli, anche se questi possono essere usati contro i propri simili.

E' forse un problema solo degli scienziati quello del controllo, e dell'uso a fin di bene, delle scoperte e delle invenzioni umane?

Precisiamo alcuni termini della questione. Lo scienziato vero, anzitutto, è quello che fa ricerca solo per amore della conoscenza: per scoprire qualcosa degli elegantissimi segreti della mirabile natura che ci circonda (opera certo non nostra, e, per chi crede, di un Dio infinitamente intelligente). Questo tipo di ricerca --che sempre meno viene finanziata nell'attuale mondo, sensibile quasi solo al denaro-- non può avere limiti, come non può subirli la ricerca poetica. E' invece il tecnologo che si occupa delle eventuali applicazioni dei risultati della ricerca scientifica (anche se lo stesso individuo, in quanto uomo, a un certo punto può smettere i panni dello scienziato per cambiare mestiere, e assumere quelli del tecnologo). Eventuali "colpe" dovrebbero essere attribuite, semmai, ai tecnologi. Ma il tecnologo stesso può giungere alla costruzione, al massimo, di *un unico* prototipo: una primitiva automobile a vapore, ad esempio. E' poi l'intervento del potere economico e politico a determinare la produzione, o meno, di innumerevoli copie del prototipo, e a migliorarlo. Il potere da controllare, pertanto, è quello economico e politico, che spontaneamente tende a ispirarsi al tornaconto, per conseguire il quale alcuni giungono a scatenare o appoggiare guerre economiche e guerre vere. E' ovvio pertanto che questo controllo non può essere demandato alle povere forze degli scienziati, e neppure a quelle dei soli tecnologi: ma esso è compito e dovere di *tutti* i cittadini.

Potremmo interpretare il messaggio di Sciascia nel senso che *anche* gli scienziati devono porsi i problemi che *tutti noi* dobbiamo porci.

Sciascia ci ricorda in tal modo la responsabilità che noi tutti abbiamo di fronte all'uso che si fa, nel bene e nel male, delle conquiste del "progresso". E bisogna stare attenti e avvertiti: perché parecchi eredi dell'antico capo tribù, che ora potremmo riconoscere, ad esempio, in alcuni controllori delle grandi società finanziarie internazionali, e anche di varie multinazionali, hanno approfittato della nostra atavica tendenza a colpevolizzare un capro espiatorio, o magari lo stregone della tribù, e si sono riparati dietro la scusa delle necessità del "progresso" e dietro un paravento di accuse ai "moderni stregoni", che hanno contribuito a identificare con gli scienziati… Qualsiasi danno ambientale o eccessivo sfruttamento della natura cominciò così ad essere attribuito alle inevitabili esigenze del progresso associato alla evoluzione della scienza. La cosa prese piede, contribuendo, temiamo, ad avviare il nostro Paese verso un Secondo Medioevo: ma chiaramente non è veritiera.

Abbiamo visto come in realtà non sia soltanto lo scienziato, o non sia soprattutto lo scienziato, ad avere le responsabilità di cui stiamo parlando. E allora, ancora una volta, come possiamo interpretare il messaggio di letterati come Sciascia, che, nel caso di quest'ultimo, lo toccava al punto da fargli attribuire tanta importanza al proprio libro su Majorana? Poniamoci una domanda, questa volta di tipo scientifico-biologico: come mai l'uomo, fra tutti gli animali, è quello che apparentemente è il più feroce coi propri simili? Perché li attacca e tortura, mentre la maggior parte degli animali non si comporta in tal modo? Una ragione biologica c'è; ed è la seguente. Gli animali che nascono con mezzi di offesa scadenti e deboli (come gli uomini, con i loro poveri denti, e unghie) non ricevono in dono dalla natura l'istinto del "cavalierismo" verso il prossimo; mentre gli animali dotati di mezzi di offesa potenti –come lupi o tigri-- posseggono di necessità l'istinto del rispetto intraspecifico: altrimenti la loro specie si sarebbe già estinta! Gli agnelli, per esempio, non hanno denti poderosi, non hanno artigli, e quindi la natura non li ha istintivamente dotati di rispetto per i propri simili; tanto che, trovandosi due agnelli sul ciglio di un burrone, può ben avvenire che uno spinga l'altro giù dal dirupo. Analogamente per due piccioni: essi pure posseggono deboli mezzi di offesa; quando eseguono le loro battaglie mimiche per conquistare il predominio su un territorio, ad un certo punto uno dei due si riconosce perdente, e se ne vola via: e basta. Ma prendete due piccioni maschi e metteteli in una unica gabbia: il vincitore torturerà a morte il perdente... Quando invece sono due lupi a recitare la mimica della loro battaglia (una mimica dalla quale, tra parentesi, nasce la nostra danza tra uomo e donna) onde conquistare il predominio sul branco, a un certo punto uno dei due lupi riconosce la propria inferiorità: questi allora si arrende, e offre il collo, esponendo la giugulare, al vincitore. Il vincitore, nonostante dimostri una gran voglia di azzannare il soccombente, in realtà è costretto dall'istinto a comportarsi da cavaliere: il primo si arrende e il secondo invariabilmente accetta la sua resa e gli risparmia la vita.

Noi uomini non abbiamo ricevuto *in dono* dalla natura l'istinto di rispettare i nostri simili. Però abbiamo poi costruito coltelli, fucili, le bombe atomiche... Cosa occorre allora? Che il rispetto dei nostri simili ce lo dobbiamo guadagnare con ogni sforzo verso una maturazione morale, che deve crescere man mano che le nostre capacità di offesa artificiali aumentano. Si potrebbe dire che Iddio non ci ha voluto fare un tale tipo di regalo, affinchè ce lo si dovesse guadagnare, liberamente, con lo sviluppo della nostra coscienza morale (a quanto pare, a Dio interessa più la libertà che non l'obbligo a comportarci bene). Vogliamo pensare che questo sia il grande messaggio di letterati e scrittori come Sciascia. Noi ora abbiamo in mano mezzi potentissimi, come aerei da guerra, bombe micidiali, armi chimiche e batteriologiche; e bombe all'idrogeno: oggi basta schiacciare un bottone per uccidere un milione di uomini. *Quindi* abbiamo l'obbligo di sforzarci –per la sopravvivenza della nostra specie— al fine di conseguire un'alta maturazione morale. Maturazione che deve certamente aver luogo nei fisici, negli scienziati, ora diremmo nei biologi, ma che dobbiamo avere tutti, perché è un compito che l'intera l'umanità deve affrontare: solo l'unione di intenti di tutti i cittadini del mondo può imporre ai veri Potenti di perseguire fini di pace.

Per concludere, se lo spazio ce lo concede, una parola per i giovani. Vorremmo ricordare a chi è ancora studente che la preparazione più importante che dobbiamo chiedere alla scuola per la vita è l'attuazione delle nostre potenzialità ereditarie, lo sviluppo delle nostre facoltà morali, intellettuali e cognitive: più che la preparazione a svolgere uno specifico mestiere. Dopo avere conosciuto alcune fette di mondo, ci sentiamo di asserire che la cultura di base che abbiamo (avevamo?) in Italia, e più in generale in Europa, è probabilmente insostituibile. Quando conoscerete altri Paesi, vi renderete conto quanto la mancanza di una forte cultura di base sia deleteria; è importante «perdere tempo» nello studio: sono proprio le cose che apparentemente non servono a nulla, come la letteratura, la poesia, la filosofia, la scienza intesa come conoscenza del mondo, che formano la mente. C'è chi ora esprime seri timori: ad esempio davanti al pericolo che molte delle nostre nazioni divengano troppo organizzate e dominanti, fino a che prendano il sopravvento i burocrati. Da svariati economisti, di vista miope, che divengono consulenti privilegiati di ministri e capi d'industria in ogni settore, a direttori-manager che hanno il compito di trasformare in aziende financo ospedali, università, enti di ricerca scientifica… Tale eccessiva burocratizzazione la si può combattere solo tenendo presente il bene comune, e vivo l'amore per il pensiero indipendente e per la cultura. Il pensiero, la meditazione, la preoccupazione per i nostri simili e per il mondo ci rendono liberi e maturi; molto più degli studi che ci vogliano insegnare *solo* a divenire ingranaggi utili allo sviluppo economico. Non tutti concorderanno con noi, a questo punto, ma riteniamo, per fare un ulteriore esempio, che occorrerebbe premere affinché venga attuato non solo il diritto allo studio, ma anche il diritto alla ricerca scientifica in senso lato; e, non meno, che bisognerebbe limitare l'istituzione di ricerche scientifiche *finalizzate,* ovvero di finanziamenti privilegiati per investigazioni da cui ci si aspettano risultati concreti a breve; perché le grandi innovazioni non scaturiscono mai da ciò che già che si conosce e si può prevedere. E intendiamo riferirci non soltanto al settore delle scienze esatte (che poi esatte non sono), ma pure a quello delle scienze letterarie e morali.

Concludiamo facendo un passo indietro, e ritornando alla circostanza che anche Sciascia ebbe a intuire che Majorana probabilmente si ritirò dalla vita consueta, senza suicidarsi; e in effetti Ettore lasciò la famiglia, e il gruppo dei fisici con cui lavorava, con in tasca il passaporto e almeno una quindicina, o ventina, di migliaia di dollari. Questa ipotesi di fuga e non suicidio ci trova consenzienti, anche in base a documenti successivamente ritrovati: i quali suggeriscono come Ettore si sia probabilmente ritirato in un luogo appartato. E' stato suggerito che il Majorana fosse un genio non solo nella fisica, ma anche in altri campi della cultura o sensibilità umana: e che abbia trovato qualcosa di più importante da fare che non adattarsi ai successi di una vita accademica e del suo genialissimo pensiero... A tale proposito appoggiamoci ad alcune parole scritteci da una letterata che opera a San Paolo del Brasile, l'italiana Aurora Bernardini: «L'ipotesi credibile e fondamentata di una sopravvivenza del Majorana è non solo più generosa, ma più rivoluzionaria o, almeno, più progressista del comodistico suicidio. Scartando a pié pari il luogo comune, secondo il quale il genio dei fisici è precoce e di vita breve, o che un fisico può avere un grande talento nel suo ambito ed essere un imbecille nel resto, stando a quanto si sa di Majorana non rimane che credere che in lui la genialità abbia anticipato la scoperta della sua verità, o della verità tout court che Ivan Iljic di Tolstoi scopre solo prima di morire. Quali sono i momenti veramente vivi della vita? Ognuno ha la sua risposta, quasi sempre in ritardo. Majorana l'avrebbe avuta prima. Sarebbe molto utile, per l'odierna umanità, il suo legato in proposito. Forse ancora più utile che il suo legato in quanto fisico».

Il fascino della vita di Majorana, e il suo insegnamento umano, sono forse questi: ovvero, quelli di uno scienziato che scopre, ad un certo punto, come più importante del ricevere dei premi Nobel sia il riuscire semplicemente a *vivere*: come ciascuno di noi; rendendo così, chissà?, gloria alla Mente suprema che gli rivelava la bellezza delle leggi che reggono il creato.

(fine dall'Appendice)

\* \* \* \* \* \* \* \* \*

# *Ringraziamenti:*

# *BIBLIOGRAFIA citata, o essenziale:*

(xiv) E. Recami, "Einstein e il rinnovamento delle scienze", *Civiltà delle Macchine* (Roma) **26** (1979) 109-112, issue no.6: arXiv:0709.2758 [physics.hist-ph].

(xv) M. Farinella: *L'Ora* (Palermo; 22 e 23 Lug. 1975); G.C. Graziosi, *Domenica del Corriere* (Milano, 28 Nov. 1972)

(xvi) C. Fontanelli: "Il caso Ettore Majorana: Aspetti storici e filosofici", Tesi di laurea, relatore A. Pagnini (Fac. Lett. Filos., Univ. di Firenze; Firenze, 1999); E. Recami, *La Stampa* (Torino; 1 giugno e 29 Giu. 1975); *Corriere della Sera* (Milano; 19 Ott. 1982 e 13 Dic. 1983).

## *Nota:*

**(\*)** Rileggiamo alcune affermazioni di Sciascia, e di Amaldi. In una lettera del 3 dicembre del '75 Amaldi ci disse: «Credo che avrà anche visto l'intervista di Sciascia sul *Giornale di Sicilia* del 9 novembre del '75. Qui finalmente viene fuori la vera posizione di Sciascia, ossia quella classica in Italia di Croce, di Gentile: la scienza non fa parte della cultura». E Leonardo Sciascia pochi giorni dopo, il 9 dicembre '75, ci scrisse, riferendosi al suo lungo articolo inviato a *La Stampa* e apparso il 24 dicembre del 1975, commentando: «Naturalmente questa è l'ultima volta che scendo in polemica, ed è il caso di dire scendo perché la polemica di Amaldi è piuttosto bassa». In un'altra lettera, il 27 gennaio '76, Leonardo aggiunse infine una interessante affermazione, che abbiamo già visto altrove: «Voglio soltanto fare presente che per me l'espressione "rifiuto della scienza" vale "rifiuto della scienza a un certo punto di fronte a certe ricerche, a certe scoperte". E cioè rifiuto da parte degli scienziati stessi».